# Computational modeling of extreme wildland fire events: a synthesis of scientific understanding with applications to forecasting, land management, and firefighter safety

Janice L. Coen[a,b]   W. Schroeder[c]   S Conway[d] L Tarnay[e]

[a] National Center for Atmospheric Research, Boulder, Colorado

[b] Corresponding author. janicec@ucar.edu

[c] NOAA/NESDIS/OSPO/SPSD, College Park, MD

[d] Conway Conservation Group, Incline Village, NV

[e] USDA Forest Service, Region 5 Remote Sensing Laboratory, McClellan, CA

**Abstract**

The understanding and prediction of large wildland fire events around the world is a growing interdisciplinary research area advanced rapidly by development and use of computational models. Recent models bidirectionally couple computational fluid dynamics models including weather prediction models with modules containing algorithms representing fire spread and heat release, simulating fire-atmosphere interactions across scales spanning three orders of magnitude. Integrated with weather data and airborne and satellite remote sensing data on wildland fuels and active fire detection, modern coupled weather-fire modeling systems are being used to solve current science problems. Compared to legacy tools, these dynamic computational modeling systems increase cost and complexity but have produced breakthrough insights notably into the mechanisms underlying extreme wildfire events such as fine-scale extreme winds associated with interruptions of the electricity grid and have been configured to forecast a fire's growth, expanding our ability to anticipate how they will unfold. We synthesize case studies of recent extreme events, expanding applications, and the challenges and limitations in our remote sensing systems, fire prediction tools, and meteorological models that add to wildfires' mystery and apparent unpredictability.

**1. Introduction**

Wildfire occurrence is punctuated by extreme events - only 3-5% exceed 100 ha in size [1] - yet the largest 1% of fires account for 80-96% of area burned [2,3]. In particular, communities around the world are at risk from large, severe wildfires that threaten people directly and create life-threatening impacts on air quality, water availability and quality, and soil integrity. A new term 'megafire' [4] was coined following the devastating 2000 U.S. wildfire season to reflect a perception that some wildfires in the U.S., and perhaps worldwide, were reaching new levels in either size, impact on society, or in severity at least in part due to a century of aggressive fire suppression. Indeed, the number of large fires has been increasing in western states[5]. Statistical analyses over broad regions indicate relationships between regional acres burned and conditions intuitively favorable for fire [5,6]. These perceptions have led to widespread attribution of large fires in total and -- more speculatively -- individual fires to climate change and fuel accumulation.

Meanwhile, physically based, mechanistic studies investigating the dynamics producing large, intense fires, confirming such speculations, have been lacking. Much progress has been made in the broad, interdisciplinary research area of wildland fire science since the 2000 wildfires, particularly in the ecological precursors to and impacts of wildfires on living systems, though fewer resources have been allocated to study the physics of fire itself. That fire behavior research has been concentrated on a sequence of instrumented small, low intensity prescribed fire field experiments the results of which have had limited applicability to large, high intensity events. Two-dimensional fire models that are based on kinematic relationships between fire behavior and environmental parameters are still broadly used for research, operations, and planning. Though these simpler tools have had wide use and utility in modeling fire spread in simple conditions, often through ad hoc calibration of inputs to obtain observed growth rates, they have been unable to represent complex mountain airflows, transient phenomena, and wildfire events that evolve through dynamic fire-atmosphere interactions - characteristics of the most destructive, highest impact events.

More specifically, the most extreme events frequently lie at an end of a spectrum. At one end, fire growth is driven by strong ambient winds, characterized by practitioners as a "wind-driven" event, while at the other end of the spectrum where ambient winds are generally weak, fire growth amplifies due to an internal feedback loop in which winds sometimes ten times stronger than ambient are generated by the heat released by the fire itself, commonly referred to as a "plume-driven" event. In such events, media commonly quote fire managers who describe fire behavior as no longer being simple or intuitive but beyond what simple operational models can predict.  Meanwhile, forests themselves are being shaped by drought, human settlement, silviculture, previous fires, blowdowns, and other disturbances. Fires in this environment are forged by intricate mountain winds, including winds amplified by the fire itself.  Thus, the complex fire environment, particularly in forested mountainous areas, exceeds current understanding and operational tools, making fires appear unpredictable, increasing uncertainty, and inhibiting the use of prescribed fires that might mitigate the hazard but carry their own uncertainties and risks.

Scientific advances such as coupled numerical weather prediction-wildland fire computational models that capture fire-atmosphere interactions and remote sensing of fires and fuels including quantitative measurements of vegetation structure are transforming fire science. These coupled models have led to better understanding of fire events and phenomena, have reproduced key features during the unfolding of large fire events (e.g., [7]), begun painting a more nuanced picture of how fire environment factors such as fuel amount and condition affect fire behavior (e.g., [8]), and, integrated with satellite active fire detection data, serve as a paradigm for next-generation fire growth forecasting systems (e.g., [9]). Coupled model simulations are broaching operational forecast use but introduce their own uncertainties associated with numerical weather prediction, namely, the decrease in their skill with the length of predicted time, limits to predictability that are particularly severe at finer scales, and ambiguities in the interpretation of fire detection and weather data that might be used to assess simulation fidelity.

Here, we synthesize research using computational models that showed how complex atmospheric flows interact with wildland fire behavior and how recent case studies of large wildfires (discussed fires are listed in Table 1) have improved understanding of transient, hazardous fire behavior and illuminated some under-recognized, difficult to understand weather-fire behavior combinations that are not well represented by current two-dimensional, kinematic models. Examples of these factors include the impact of gust fronts and shifting winds on fire behavior, fire-induced winds, large fire whirls, transient plume behavior, coastal airflows, and complex topographic airflow effects. In addition, we discuss challenges and limitations in our computational models, remote sensing systems, and underlying meteorological models that add to wildfires' mystery and apparent unpredictability. As presented, Section 2 describes the models used in wildland fire simulations, including the considerations associated with developing and applying models that shift land management from widely used laptop-based empirical projections to high performance computing-based expert systems. It also describes the expanding pool of fire detection and mapping data available to initiate and evaluate simulations and surface weather station networks tasked with revealing complex mountain airflows. Section 3 describes application of the new generation of models to scientific investigations of the mechanisms driving extreme fires, practical uses in fire management and mitigation, and distilling knowledge about complex, transient fire phenomena to improve wildland firefighter safety. Differences in the predictions from what should be similar systems are noted. Section 4 draws together the knowledge, technical, and safety advances brought forth by the infusion of computational modeling into an area traditionally relying on empirical learning and calls out areas needing further advancement.

**2. Tools and Methods**

As the complexities of wildfire behavior became more recognized, fire modeling tools evolved from simple algorithms to computational models. This section discusses new modeling tools, data sources that may be used to initialize and assess them, and the constraints and limits that accompany these more complex models.

*2.1    Wildland fire modeling*

Large wildland fires are dynamic, complex phenomena. They encounter a wide range of fuels, terrain, and weather - often during one event - and produce extreme behaviors such as fire whirls, blow-ups, 100-m long bursts of flame shooting ahead of the fire line, fire winds that are ten times stronger than ambient wind speeds, deep pyrocumulus clouds, and firestorms – all resulting from the interactions between a fire and its environment. Despite emergency alert systems and ever improving weather forecasts, communities may be tragically unprepared for complex and explosive fire behavior, for example the 2018 Carr fire firenado, which burned into Redding, CA, and produced 230 km h$^{-1}$ (143 mph) winds. Indeed, wildfires are often described as 'unpredictable' because human intelligence and current generation fire prediction models cannot integrate all the interacting atmospheric dynamics to anticipate when weather will combine with topography and fuels to dramatically amplify fire behavior.

Current fire modeling tools used in operations [10,11,12] are kinematic, i.e. they relate fire motions to environmental factors without considering the forces that cause the motion and do not model transient behavior and phenomena. Using weather station data, diagnostic wind tools, or gridded weather forecast output, and ad hoc calibration of inputs (e.g. [13]) that conceal model limitations to match observed growth rates, these tools have provided estimates of fire growth rate and direction in some simple conditions.

The underlying semi-empirical formulae are algebraic relations that express the flaming front rate of spread as the theoretically based rate of spread in zero wind conditions on flat ground, adjusted by empirical factors determined from laboratory and field experiments dependent on the wind speed at midflame height and terrain slope. These relations have been criticized for assumptions regarding their representation of combustion and inadequacies when tested against real fires using nearby winds. However, the more coupled sophisticated models described in 2.1.1 and 2.1.2 that use these semi-empirical relationships as parameterizations at large scales in dynamic models have shown they do capture key sensitivities - how fire rate of spread responds to environmental factors of fuel, wind, and slope - and can represent fire behavior, phenomena, and event evolution well when the temporal and spatial variability of wind including fire-induced winds from the fire are calculated accurately at microscales (100s of meters). This can be scientifically and computationally challenging, but is within the grasp of some computational modeling systems.

### 2.1.1 Computational aspects of coupled atmosphere-fire modeling

A newer generation of models couple a fire behavior or combustion physics module with a three-dimensional computational fluid dynamics (CFD) model. These range from coupled systems designed for small scales (grid spacing 1 m or less) (e.g., FIRETEC [14] and WFDS [15]), where the CFD model represents airflow in a small domain (~ 1 km$^3$) and employs detailed parameterization of combustion of wildland fuels to systems designed for large-scale fires. The latter (CAWFE [16,17], WRF-based systems such as WRF-Fire [18] and derivatives, and the Unified Model [19]) are based on Numerical Weather Prediction (NWP) CFD models, include additional variables and processes representing phase changes, large-scale weather patterns, and grid refinement over nested domains potentially spanning the globe and parameterize fire processes, which occur at scales much finer than the resolution typically used to model weather(thousands to tens of thousands of meters grid spacing). These NWP-based coupled weather-fire models do not represent combustion details but instead see the fire as a spatially and temporally varying heat and moisture source, an approach that results (e.g. [9]) have shown can sufficiently capture fire event dynamics using one to a few hundred meters grid spacing by capturing convective-scale winds induced by topography and the fire itself. Each type is best suited for different types of fire modeling problems [20]. The sets of equations composing each system consist of fundamental rules governing the physics of fluids - the second law of thermodynamics, the ideal gas law, a statement of mass conversation, and equations describing the rate of change of momentum and scalars including mass concentrations of atmospheric water phases [21]. Consequently, this coupling achieved through computational

models - as the heat and water vapor released by the fire create a buoyant updraft over the air, drawing air in at its base to replace air rising in the column and altering the winds directing the fire - allows dynamic feedbacks between the fire and the fluid medium in which it occurs, creating fire phenomena such as firenadoes, allows effects from one environmental factor to modulate or amplify another, and creates the observed upscaling as individual fire plumes grow into towering pyrocumulonimbus storms penetrating the stratosphere. Accompanying side effects include more complexity, the requirement for computational weather model knowledge, greater computational resources, and the decrease in skill with time and limits to predictability that accompany NWP models. The last issue is an important constraint on coupled weather-fire simulations because the rapid error growth associated with the required one to a few hundred meters grid spacing limits simulations to one to two days, or less in meteorological or fire conditions where uncertainty increases rapidly [20].

    Key to the computational solution for fire growth in a coupled system is the underlying NWP model. Numerical weather prediction models are often designed for simulation of specific scales of atmospheric motion, from coarse synoptic scales, through mesoscale simulations (with the finest horizontal grid spacing of 2-20 km), convective scales (hundreds of meter - a few kilometers grid spacing), and even finer turbulence scales. NWP models have been used for decades to simulate weather phenomena that are also linked to critical wildland fire behavior - atmospheric convection, cloud production, outflows, gust fronts, and microbursts, mountain circulations, and downslope windstorms. Though NWP models may solve the same set of equations, different approximations, solution methods, and discretization can produce solutions with different properties, with impacts on simulations of fire behavior (discussed in [20]). For example, the WRF model produces an atmospheric energy spectrum that adheres closely to the

expected natural spectra across synoptic and mesoscales [22], but components included to reduce numerical divergence severely dissipate fine-scale motions and smooth fine-scale gradients. Studies have shown this has led to underestimates in peak wind extrema in wind-driven fire events [23], smoothed gradients across interfaces precluding WRF from producing gravity wave overturning (discussed in Coen 2018) - a primary mechanism for producing strong downslope winds driving fire events such as the 2012 High Park Fire [24] and shear instabilities [25], and smoothing gradients underlying simulation of the full development of fire phenomena such as fire whirls [20,26].

### 2.1.2 CAWFE

The CAWFE® modeling system (derived from Coupled Atmosphere-Wildland Fire Environment) combines a numerical weather prediction model with a module describing fire behavior and fuel consumption. The weather and fire components are coupled so that fluxes of heat and water vapor from the fire alter the state of the atmosphere, notably producing fire winds, and the evolving atmospheric state affects fire behavior. CAWFE is built upon the three-dimensional time-dependent meteorological Clark-Hall model [27,28], the solution methods and options of which were designed to excel at very fine scales (~100 m grid spacing) in extremely complex terrain. Vertically stretched terrain-following coordinates allow detailed simulation of airflow at horizontal resolutions of tens of kilometers while telescoping in to focus at hundreds of meters in the area of a fire. The outer of several interactive, nested modeling domains are initialized from and boundary conditions are updated with gridded atmospheric analyses or large-scale model forecasts.

CAWFE's fire module [16,17] contains semi-empirical algorithms that parameterize subgrid-scale fire processes. One algorithm predicts how fast the flaming front of a surface fire will spread in wildland fuel complexes [29], as a function of terrain, fuel properties and moisture state, and wind, which may be altered by the fire. The limits to direct use of Rothermel's semi-empirical relationship [29] for fire spread are well known but when used with fire-induced winds - the importance of which was not recognized until the 1990s -- it has proven very successful, likely because the response to increases in wind speed - the dominant factor - and slope are based on empirically derived relationships. Additional algorithms calculate how rapidly fuel complexes behind the flaming front are consumed depending on their size composition [30], whether the surface fire is intense enough to ignite and transition to the canopy, if present, and how fast the subsequent crown fire would spread [31]. As fuel is consumed by the surface and crown fires, sensible and latent heat is released as well as smoke particulates (or other chemical species) via an emission factor. A radiation treatment distributes sensible and latent heat fluxes and particulates from the fire into the lowest atmospheric grid volumes.

CAWFE has been used to simulate fire spread and phenomena from prescribed to landscape-scale wildfires. Its reproduction of fire behavior and expansion was evaluated in over three dozen large wildfire events. As a coupled computational modeling system, it provides a physically-based, internally consistent link between fuels, weather, terrain, and the fire behavior, intensity, and effects that result. For example, spatial maps of simulated heat flux were used to investigate if remotely sensed fuel properties could improve predictions of burn severity during studies of the King Fire [8,32]. Simulations have reproduced transient fire phenomena such as horizontal roll vortices (HRVs), splitting or merging of the fire line, fire whirls, rotating convective columns, multiple convective plumes along a fire line, sudden changes in fire direction, pyrocumuli, and blowups.

CAWFE simulations ingest data on terrain elevation, gridded weather, and fuel characteristics/properties, along with the fuel state, and fire location or extent. Gridded weather data is used as initial conditions and lateral boundary conditions for the outermost of several interactive, nested CAWFE modeling domains. NOAA provides regular three-dimensional model state data; this can be from either gridded atmospheric analyses for past events or large-scale forecasts for predicting the future.

Fuel data describes surface fuels, categorized into stylized fuel models according to a fire behavior fuel classification schemes (e.g., the Anderson (1982) 13 [33] or Scott and Burgan (2005) [34] 40 category systems) with which are associated typical properties for each category or custom data from the user. Additional fuel information describes the tree crowns such as canopy bulk density, base, and height, and ground fuels such as duff. Fuel mapping data for most systems is now available wall-to-wall across the U.S. but still have some issues such as classifications may be unevenly validated. The data contains errors in type and amount but these are unquantified. Dead fuel moisture is taken from reports, from RAWS data, the National Fuel Moisture Database (NFMD), or diagnosed or predicted within CAWFE from atmospheric conditions. Live fuel moisture is obtained from incident or accident reports and NFMD.

*2.2 Fire detection and mapping*

Wildfire modeling research and decision support rests on information about fire occurrence, their extent, intensity, growth, and effects on the air, landscape, and ecosystems.

Information for input into modeling systems and validation is currently served by a disparate collection of observations from remote sensing systems with a wide range of revisit frequencies and resolutions. Fire occurrence and spatial mapping is derived from a pool of satellite and airborne sources, some (like satellite data) encompassing the globe and others performed on demand such as collection of scientific research datasets or NIROPs -- night-time observations performed on the highest priority wildfire incidents -- or research flights in support of scientific investigations. These sensors have tradeoffs between spatial and temporal characteristics, ranging from geostationary observations with updates every few minutes but low spatial resolution to very high spatial resolution sensors that return infrequently.

Multi-decadal records and products have been derived from moderate resolution sensors since the 1970s. For example, the MODIS active fire product [35] detects fire occurrence but is unable to define fire outlines except for the largest fires. Products from more recent sensors such as the I-band of the Visible and Infrared Imaging Radiometer Suite (VIIRS) [36], the first of which began returning data in 2011, have spatial resolutions that can delineate actively burning fire lines and at least twice daily observations at midlatitudes and are particularly apt to detect smaller/lower intensity fires, while enabling improved flame front delineation over larger wildfires [37]. The 2017 launch of a second instrument augmented VIIRS' biomass burning mapping capabilities. In addition to VIIRS, a fire detection algorithm was developed for Landsat-8 Operational Land Imager (OLI) 30 m resolution imagery [38] and implemented operationally at the USFS Geospatial Technology and Applications Center. Though OLI data have higher resolution, they have higher latency and repeat coverage occurs every 16 days. . That fire detection algorithm has also been expanded to ESA/Sentinel-2 Multi-Spectral Imager (MSI) data.

The Advanced Baseline Imager (ABI) on the geostationary GOES-16 satellite added new opportunities to track wildfire evolution. Despite 2 km spatial resolution, ABI "flex mode" provides full hemispheric observations every 15 min and 5 min coverage of the conterminous U.S. with a 60 s repeat imaging of up to two 1000×1000 km subsets of relevance or one subset every 30 s. Though too coarse to delineate a fire, the variation with time can be used with VIIRS data to constrain daily variability in time-sensitive fire energy radiative fluxes.

The satellite observations come from sensors with various spectral characteristics but most frequently use infrared bands as they can detect fires through smoke. Mid-wave IR products are most widely used for detection (e.g. Fig. 1). and underlie products indicating the energy release rate such as Fire Radiative Product but other IR bands also detect aspects of fires, for example, short-wave infrared (SWIR) bands (1-3 micrometers) detect active flaming regions while long-wave infrared (LWIR) bands (8-14 micrometers) record the hot ground under and behind the flaming front's passage.

Meanwhile, the private sector has actively been developing and deploying space-based commercial imaging systems that complement those government systems. Though many, including DigitalGlobe and emerging small sat constellation company Planet, have targeted ultra-high resolution visible imagery, in August 2014, DigitalGlobe added the WorldView-3 satellite, which makes 7.5 m pixel imagery in 8 short wave IR (SWIR) bands (1.2 - 2.4 micrometers) commercially available, with a revisit frequency of under a day. These have been shown to detect actively burning regions of wildfires however, at nadir, a swath width is 13.1 km, so adjacent swaths collected at different times may need to be mosaicked to image an entire large fire, creating discontinuities in fire data. Additional private sector initiatives aim to provide high-resolution thermal imagery achieving high revisit rates through a constellation of sensors.

Other data sources include airborne mapping data from the USDA Forest Service National Infrared Operations (NIROPs) where available, and incident maps. Ignition data is obtained from incident reports or VIIRS' first detection. Airborne data has been collected during research experiments with airborne medium wave infrared sensors [39] and multispectral systems such as the USFS's FireMAPPER [7,40]. Operationally, NIROPs is dispatched to high priority fires at night and Colorado's Multi-Mission Aircraft may map fire activity with surveillance-band infrared imagery. Additional sensors are becoming more common. UASs have been used in prescribed fires [41] and demonstrated over wildfires [42] though their broader use is still restricted.

Mapping and interpreting fire behavior and extent in some fires, particularly some with the highest impact, remains challenging. Current fire mapping data is problematic for studying fires at times in between observations, for fast moving fires that may have ignited and reached tens of thousands of acres by the time of first observation (such as recent northern California wind-driven events or High Plains range fires), and for representing the important diurnal variability in fire growth that is not captured in twice-daily observations. Rapidly moving fires in the southwest in sparse fuels over hard soils may not retain enough heat signature to be detected. Also, associated variables needed for analysis but which are not easily observed (such as the near instantaneous rate of spread versus a many-hour or daily average between consecutive observations) are not recorded. Post-processing of high-speed airborne infrared data [40] has been used to extract rate of spread in naturally burning fires.

In addition, in fires producing copious numbers of embers, difficulty remains in remotely distinguishing between the location of the flaming front and spot fires that have landed, as well as between landed and airborne ember storms. Additional uncertainty in fire extent on the ground can occasionally be introduced by fire plumes filled with $CO_2$, condensed water, or embers when viewed obliquely.

*2.3     Surface weather station networks*

The U.S. weather station network consists of National Weather Service (NWS) stations, Remote Automated Weather Stations (RAWS) maintained by land management agencies, and smaller-scale networks including those operated by Oklahoma, New York state, and, recently, expanding networks placed by utility companies in California. Standards for siting stations [43] consider general requirements (meteorological variables, location, sampling frequency, and timing) for installing stations in view of the application areas for which data is needed, for example, NWS stations are often co-located with airports and RAWS stations are located where they can monitor fire danger. Siting is constrained by land physical characteristics (e.g. avoiding where trees or terrain may disrupt airflows) and land ownership and access, sometimes limiting stations to non-optimal siting at accessible locations where land is owned by the public or a cooperative private entity. Set amidst forested terrain, stations may experience blocking from certain wind sectors. The distance between stations, especially in the western U.S., remote areas, and wilderness where it may be tens of kilometers, is problematic for wildfire applications. Neither network is targeted for detecting microscale variability in complex terrain or where winds may disproportionately impact the electrical grid.

In conjunction with national and utility-operated forecast models based upon the WRF mesoscale model, this intelligence network left numerous weakness with respect to diagnosing and predicting conditions that may impact fires. These weaknesses are (1) mesonets do not resolve fine-scale wind circulations or extrema that occur between stations, (2) mesonets are limited to near-surface observations, (3) plume-driven fires, which are often among California's most destructive, often occur during weak wind conditions and arise from undetected dynamic interactions between fire-induced winds, fuels, and topography, and (4) the WRF model's numerical properties are diffusive of fine-scale gradients [20,22], which, in simulations of recent wind-driven California fire events, has smoothed out extrema and led to underestimates of 10 m s$^{-1}$ in peak winds, e.g. [23].

Current activities and the California utilities' wildfire mitigation plans include dramatic supplementation of the existing weather station network. These promise greater situational awareness and better anticipation of weather events that, when interacting with the grid, may ignite fires or spread fires that impact the utility infrastructure and other assets at risk. Remaining issues relate to validation of weather prediction models and the use of surface weather station data to do so. The user community implicitly assumes that weather station data resolves the essential nature of airflow, building upon the longstanding assumption that smaller-scale motions are weak and unimportant, and that, despite the terrain variability, weather model simulations (where grid point predictions represent averages over the grid volume and time step) with effective resolutions of several kilometers can be directly compared against station measurements; agreement is interpreted as success while disagreement is attributed to "gusts" not represented by the spatial or temporal resolution of the model. Direct comparison between surface measurements and model predictions is additionally complicated in that some vertical displacement is required between the model grid point and measurement height. It is assumed that standard profiles for how the along-surface wind speed varies with height in the atmospheric surface layer (a log profile) or atmospheric boundary layer (a power law), however, none of the assumptions underlying derivation of those profiles are satisfied in these airflows (that the airflow is stationary, the terrain is flat, and that the air is neutrally stratified), thus, it is expected that the vertical variation of wind speed can vary substantially from these assumed profiles.

Subsequent studies with models suitable for finer-scale simulations have suggested the presence of various types of coherent, dynamic microscale phenomena largely unexplored in the scientific literature; these phenomena create local extrema with peaks 5-10 m s$^{-1}$ higher and time series of wind speeds produce oscillations of 4-13 minutes [7,23], a frequency noted in station data reported at high frequency, the mechanism for which has not been explained. Subsequent targeted siting of utilities' surface stations in high wind zones have produced measurements of higher gusts, supplementing photographic evidence of wind damage consistent with EF1 tornado-strength winds in suspected wind zones [M. Mehle, pers. comm.].

NWS and RAWS stations are frequently relied upon as evaluation for NWP model simulations, but the direct comparison between the station measurement, presented as a several minute mean and peak gust, and the volume- and time step-averaged mean of a model is challenging, particularly when the near surface airflow varies greatly in space and time. Direct comparison would make sense in uniform flow in more uniform terrain, but during recent events, a view of observations from the NWS and RAWS stations failed to convey the extreme nature of events, however, as noted, new stations, particularly in northern California where some have been sited in known wind zones and in previous wind-driven fire boundaries, higher gusts reaching 45 m s$^{-1}$ have been reported.

## 3. Applications

In this section, we synthesize some recent studies to give insight on some overarching fire research issues and the new generation of computational weather - wildland fire models being used to address them, such as what causes distinct types of outlier fires. We also discuss applications, such as identifying situations with potential for large fire growth, fire growth forecasting, and how studies of transient fire behavior are being used to improve firefighter safety.

In typical use, a coupled weather-fire simulation is initialized and the boundary conditions later updated with large-scale atmospheric analyses (for past events) or forecasts (for future events). The fire is initialized either at its known ignition time and location or, for later periods in the fire, initialized in progress using a fire perimeter. From then, 12-36 h periods of interactive weather and fire growth can be simulated well until fidelity has degraded due to the accumulation of errors. In previous applications, CAWFE has shown that provided winds (1) are well reproduced at a few hundred meters resolution and (2) include feedbacks from fire on atmospheric dynamics, much of the unfolding of large events including distinctive shapes of the expanding perimeter and fire phenomena can be reproduced. Extensions include the connection to model fire effects, such as to atmospheric chemistry modules for modeling air quality effects [44] or the calculation of burn severity [8,32].

Some studies have used idealized studies to address basic aspects of fire behavior, e.g. [45], and sensitivities [18]. Numerous studies target model validation, e.g [46], while others sought understanding for aspects of fire behavior, e.g. [7], or overall event growth, e.g. [24]. There remains misunderstanding about why a particular fire became large and what factors in the fire environment were important; few events have been scrutinized with physically-based studies.

*3.1    Investigating physical mechanisms behind extreme fires*

The most extreme events frequently lie at either end of a spectrum. At one end, fire growth is driven by strong ambient winds, characterized by practitioners as a "wind-driven" event, while at the other end of the spectrum, fire growth amplifies due to an internal feedback loop in which winds are generated by the heat released by the fire itself, commonly referred to as a "plume-driven" event. Coupled modeling studies, as described below, have allowed investigators to deconstruct events, identifying the physical mechanisms that lead to exceptional fires, and interpret distinct periods and regions of each type of behavior within vast, complex, long-lasting events.

### 3.1.1    Plume-driven megafires

Public alerts such as Red Flag Warnings address the potential for large wildfires in terms of external environmental factors, emphasizing extreme conditions. These overlook the potential for rapid growth in unexceptional conditions, where ignition in a fortuitous location finds fuel and terrain ingredients for rapid growth even in weak winds. In addition, despite the perception that long-term impacts such as drought govern fire behavior, research [47] suggests short-term weather has more effect. Fire growth is generally episodic, tied to few day period of favorable weather conditions. For example, Peterson et al. (2015) [48] found the two days of rapid growth during the 2013 Rim Megafire were a disastrous consequence of southwesterly winds nudging an ignition at the base of inclined canyons and a second day when fortuitously aligned southwesterly winds fanned its rapid growth further up into the bowl, creating a pyrocumulus cloud [49] that may have enhanced fire growth. (They did not have means to assess the magnitude of fire induced winds.)

Winds within a wildfire may greatly exceed ambient conditions, in some cases by a factor of ten [50] and be of tornadic strength. Current kinematic operational tools poorly represent plume-driven fires because fire spread is generated by internal dynamics, not ambient winds. Particularly on inclined terrain, fire-induced winds have the potential to greatly accelerate fires, an effect that may be channeled and focused by topography (i.e. the "chimney effect"). An example is the 2014 King fire, which typifies the difficulty in anticipating fire behavior in complex terrain where environmental winds are weak without computational models.

The 2014 King Fire occurred amidst a prolonged drought in the Sierra Nevada Mountains. Following several days of slow growth through rolling hills in weak southwesterlies , the fire entered the base of the Rubicon River Valley and, while nearby weather stations recorded weak to moderate winds like previous days, raced 25 km in one afternoon to the top of the canyon (Fig. 2a). CAWFE simulations [8] reproduced the rapid growth (Fig. 2b), smoke transport (Fig. 2c), and phenomena including multiple heading regions as fingers of rapid growth found favorable topographic niches to draw themselves up as the fire travelled over rolling hills approaching the canyon and a pyrocumulus cloud in the plume over the fire's heading region.

Coupled systems can explore broader questions about what factors were important in causing such an exceptional event and how they might be mitigated. For example, additional simulations used CAWFE to disentangle the mechanisms causing rapid growth. To reveal the magnitude of fire-induced winds, a second experiment without a fire was conducted, and the west-east, south-north, and up-down components of air velocity in one experiment are subtracted from the other, revealing the location and magnitude of the fire's impact on the winds. This showed fire-induced winds during the King Fire were on the order of or exceeded ambient winds [8], and, along with a fine-scale wind circulations beneath the resolution of weather station data within the Rubicon Canyon, were responsible for the unanticipated growth. Drought and fuel loading were found not to be determining factors, especially on flat terrain, but results showed that these factors could amplify their individual effects jointly when a fire was climbing inclined terrain. Thus, a broader implication is that, in such fires, fuel mitigation might be prioritized on slopes, where it could have the biggest impact in limiting fire behavior. And like many fires of this type, these factors had no opportunity to affect size, because though they might modulate fire spread rate, the extent was limited by rocky terrain.

Part of needed research includes re-examining beliefs arising from legacy kinematic models, for example, that megafires require extreme fuel loads - a fire community perception perhaps arising from kinematic models' inability to create extreme rates of spread without artificially adjusting fuel loads to extreme values. Similar assumptions are widely held that beetle killed-trees, which have lost moisture in their crowns and therefore are presumed to be more flammable, are at risk for more intense fire behavior but [24] showed the 2012 High Park megafire's growth was not enhanced by beetle kill and that through generating a more intense plume, might somewhat hinder downward spread in fires driven downslope by strong winds.

Researchers have developed growing interest in pyrocumulus clouds (e.g. [49]) because of the potential for precipitation to influence fire behavior and because deep plumes may inject smoke into the stratosphere, affecting its radiative and chemical properties. Future research could benefit from further application of computational science bridging the apparent disconnect between the general (large-scale correlations between fire areas and drought metrics) and specific events (where certain factors may only affect certain fires in certain scenarios).

### 3.1.2. Wind-driven megafires

Large, high-impact fires can occur when strong wind events overlap sufficiently dry conditions. Several of the most destructive recent U.S. wildfire events were associated with strong downslope winds in the lee of mountain ridges, where winds are variously referred to as Santa Anas in southern California, Diablo winds in northern California, chinooks (unseasonal July Colorado Front Range downslope windstorms), or the more general term föehn winds for cross barrier flow in the eastern U.S. Appalachian Mountains. In addition, numerous events may be associated with broad areas of strong winds associated with synoptic-scale events such as pre-frontal, frontal, or pre-hurricane winds driving massive range fires in the south central U.S. plains or, in the late 1800s, the Great Lakes region.

In the case of the topographically-tied events, while an underlying synoptic-scale pressure gradient driving winds from high to low pressure invariably sets up a wind event, the mechanics establishing exceptionally strong winds - determined by flow parameters including the terrain aspect ratio (height divided by half-width), wind speed, and atmospheric stability - can vary greatly between events and locations. The strength and gustiness of one of the most studied regimes, Colorado Front Range windstorms, have been explained in terms of (1) an imperfect analogy to a hydraulic jump [51] of a stream, (2) reflection of upwardly propagating atmospheric gravity waves from a critical level [52,53] where a critical level may be defined as where the wave phase speed is equal to the mean flow velocity, thus, for the case of stationary mountain waves, where the cross-barrier mean flow is zero, or (3) vertically propagating gravity waves became unstable and overturned (i.e. broke) [54,55] creating significant increases in lee slope surface winds. (An example of a CAWFE simulation of gravity wave overturning and breaking during the High Park Fire is shown in Fig. 3.) Analytical solutions exist for only a few idealized problems, thus, research in airflow over complex terrain has relied upon computational modeling.

Mountain waves and related airflows around the world have been studied with NWP models. Two that have been applied to wildfire cases are the Clark-Hall atmospheric model (the basis of CAWFE) and WRF (the basis of WRF-Fire and derivatives). Computational modeling has been used to address current issues relating to the impact of wind events on the California electrical system and subsequent fires - events that have in the past several years caused several multi-billion dollar events. Investigations have attributed the ignition of several recent fires (the 2007 Santa Ana-driven Witch and Guejito Fires in southern California, the 2017 North Bay Fires in northern California, the 2017 Thomas Fire, and the 2018 Camp Fire) to utility malfunctions that occurred during strong wind events. A primary issue was the magnitude of peak winds, as utilities are required to maintain and clear lines of vegetation to withstand winds of approximately 25 m s$^{-1}$. In 2019, utilities began to apply Public Safety Power Shutoffs (PSPSs) in areas forecasted to be at risk of strong winds that might similarly ignite and spread additional fires. A second issue has been the nature of the airflow regime itself. Because of the sparse observation network and lack of very high resolution studies of fire-related windstorms in that area, mechanisms associated with strong lee slope winds were not known.

Case studies have simulated the airflow and, in application of coupled weather-fire models, fire growth associated with several of these events. Simulations at sub-kilometer grid spacing using the mesoscale WRF model, for example, simulated the Witch Fire's acceleration down the lee slopes in San Diego County [56] the Tubbs Fire [57,58] (Fig. 4), and the Camp Fire [59]. In each case, the flow shows acceleration on lee slopes and resembles a hydraulic jump with some steepening of potential temperature surfaces, with modeled winds of 25-31 m s$^{-1}$. These show the success of mesoscale models applied at high resolution in identifying areas of strong winds, though they are very smooth, despite intricate topographical variations, and additional 'gust factors' are necessary to explain measured winds or wind damage that exceed these values. Additionally, WRF's severe dissipation of fine-scale motions has detrimental effects on reproducing sub-kilometer motions, precluding simulation of processes that require resolving fine-scale gradients, notably wave breaking, leading to underestimates in predicted wind speed [20]. Thus, these studies leave unanswered the questions: What are the mechanisms for exceptionally strong winds and where specifically might these occur?

Several studies have applied CAWFE to these wind driven events to investigate the development of unrecognized extreme microscale wind phenomena and how these made fire events unfold, and the difficulty of using observations at the time to evaluate them. These reproduced key features and timing, and showed that strong winds in the Tubbs Fire [23] (Fig. 5a) occurred on secondary hilltops, rather than tall ridges, and produced pulses of strong winds exceeding 30-40 m s$^{-1}$ with 4-13 minute pulses that appeared in high temporal resolution wind speed data. Protuberances on the southern flank, seen in active fire detection data (Fig. 5b), can be attributed to the fire drawing itself up topographic features. Further north, the Redwood Valley Fire (Fig. 6a) occurred when the pressure gradient drove air over a lower barrier in the Sierras, creating a shallow, narrow river of high speed air (Fig. 6b and 6c) that reportedly caused both ignitions of the Redwood Valley Fire (Fig. 6d) and its subsequent deadly run (Fig. 6e and 6f). The Camp Fire ignited during strong winds and overran the towns of Concow and Paradise within a few hours as seen in Landsat OLI and VIIRS data (Figs. 7a and 7c) and was reproduced in CAWFE simulations (Fig. 7b and 7d). Strong winds were present all along the Sierras, but uniquely strong in the fire area. CAWFE simulations showed that pulses of gusty winds of 30-40 m s$^{-1}$ drove the fire rapidly downslope (Fig. 8a, 8b), encountering the town of Paradise from the side within 4 hours (Fig. 7b) and also showed that the 4-13 min pulses, seen in high temporal frequency station data, were caused by a shear instability. A velocity difference across the interface between the vertically stratified fluid occurred as the near-surface river of stable air was lifted into slower moving air, crests of high momentum air retroflected (crashed backwards) and down to the surface, bringing gusts that drove the fire downslope. Significantly, WRF simulations of the weather during this event [59] produced only some supercritical flow over the upper slopes (Fig. 8d) and though VIIRS observations show the fire extent having descended the slopes and entering Paradise by late morning, similarly configured simulations of this event using WRF-based coupled weather-fire models (unpublished results, https://www.youtube.com/watch?v=qpm0nq4rhdU and https://www.youtube.com/watch?v=k2XG0CNHMEk), did not produce this shear instability, did not drive the fire down the mountain, and predicted that even after three days the Camp fire would never have reached the town of Paradise. Thus, coupled weather-fire simulations configured similarly, which differ primarily in the underlying NWP model, may give dramatically different results.

Extreme fires associated with downslope winds are not limited to the mountainous western United States and southern Europe. A downslope windstorm (with maxima reaching 39 m s$^{-1}$) overlapping severe drought drove the Chimney Tops 2 Fire in Great Smoky Mountain

National Park down into Gatlinburg, TN, killing 14 (Fig. 9a). While unusual, a CAWFE simulation (Fig. 9b) showed the airflow resembled a hydraulic jump, with strong winds descending from steep topography before sharply turning upward, marking the end of the strong wind zone, similar to that of windstorms more commonly associated with the Front Range of Colorado. The surprisingly rapid fire spread (CAWFE simulation shown in Fig. 9c) followed the localized pulsing downslope winds down the north face of the Smoky Mountains to where and when the strong wind zone ended. Again, compared to the actual growth (Fig. 10a), and CAWFE simulations (Fig. 10b), WRF-based coupled weather-fire simulations of the event [60] (Fig. 10c) produced strong winds that underestimated the event and underpredicted fire extent by a factor of 10-20, predicting the fire never approached Gatlinburg, TN.

These case studies point to a wide range of possible airflows, with the potential to produce local extrema, where the impacts of a fire ignition can lead to $1B losses. As forecast guidance, we see that different underlying computational weather models can produce starkly different results, suggesting a key area for additional computational science investigations.

Current research focusses on fire spread in the wildland urban interface and on the role of embers (e.g., [61]). Embers have been identified as a key mechanism in structure ignition. They can also play many different roles in fire spread in wildlands, either through long-distance spotting of individual embers, short/medium distance spotting ahead of the fire line where they may or may not be overrun by flaming front before can create a plume of their own that draws the original fire forward. Or, as in many events, it may be difficult to identify a 'flaming front' amidst an ember storm. While numerous studies have considered ember transport in idealized conditions or interior chambers, computational science modeling in ember transport in outdoor, landscape-scale fires is in its infancy and is complicated by terrain- and fire-modified winds.

Embers complicate validation, as it may be difficult to identify the leading edge of the fire and satellite active fire detection data, which is commonly used for validation, may be difficult to interpret [23], as detections may be airborne embers or smoke plumes rather than surface fire.

### 3.1.3 Complex events

Computational models coupling weather and fire have supported unique insight into explaining what caused complex events to unfold the way they did. Fatality reports following firefighter fatalities attempt to assemble all available information including weather and fire behavior in addition to human actions to distill Lessons Learned that improve firefighter safety. In complex situations, traditional fire behavior tools and knowledge have left investigators speculating about complex interactions between fire, weather, and terrain. This distinction is important because without comprehensive understanding of what actually produced an event, we cannot recognize and mitigate controllable factors to prevent it from happening again.

Some events most challenging to understand are complex wildfire events that develop both plume-driven and wind-driven heading regions. While Santa Ana fires, for example, are thought of as a regional phenomena, large fires may have components that act somewhat independently. For example, CAWFE studies of the Santa Ana-driven 2006 Esperanza Fire [7] showed it split into a wind-driven component running across the base of Mt. San Jacinto, while a plume-driven component drew itself apart and raced up a drainage orthogonal to the winds, killing 5 firefighters.

The massive 2017 Thomas Fire occurred in southern California during an extended, two-week long Santa Ana event. The progression map (Fig. 11) highlights how complex and mixed the factors driving fire behavior can be, showing that the first and sixth days accounted for most growth, mostly east to west, while on days in between, the fire spread laterally into topography to the north and south. Large-scale weather forecasts and VIIRS observations were used to initialize a sequence of CAWFE simulations covering a six day period in order to deconstruct the fire-driving mechanisms [632The simulations showed that both ignitions and the first period of fire growth occurred in a shallow river of extreme Santa Ana-driven flow south west down the Santa Clara River Valley (Fig. 12a-c). For several days, while Santa Anas are reported in the area, the fire then spread laterally (Fig. 13a-c), drawing itself up canyons orthogonal to the wind, where the fire was sheltered from the Santa Anas and in other places enhanced by topographically channeled winds. In some periods, it created deep plumes that rapidly pulled it up north-south aligned drainages and bowls. The Santa Ana strength waxed and waned, the areas that experienced winds varied with time, and during the second rapid growth period, the Santa Anas surfaced in areas where the fire had reached, again creating a wind driven period west along the coastal mountains. Computational modeling of this event was extremely challenging due to the vast size, the numerical challenges of reproducing very fast airflow through extremely sharp, steep, highly varying terrain, compounding errors, and urban and firefighting effects on fire behavior.

*3.2    Fire Management and Mitigation Applications*

The most obvious use of models such as CAWFE is in anticipating where and how fast a fire will grow in the future. We discuss this use and others with potentially greater impact - identifying underestimated locations and conditions that, should a fire start, could, through dynamic interactions, lead to rapid fire growth and testing mitigation strategies in complex landscapes.

### 3.2.1 Fire growth forecasting

The coupling of NWP models to fire behavior modules allowed researchers to simulate the evolution of fire events with much greater fidelity and introduced the possibility of their use as forecasting systems operationally for fire management. However, forecasting still had several challenges. First, the skill in NWP forecasts decreases with time, such that short-term forecasts may be accurate, but as the length of the forecast increases, errors increase; in fine-scale forecasts such as are used here, the forecast may only be useful for 1-3 days. In contrast, a fire event might continue for several weeks, thus, a single simulation could not accurately encompass a fire event. Secondly, some aspects of fire growth are inherently unpredictable. For example, the spotting process whereby embers lofted ahead of a fire where they may ignite another fire, contains several steps what one may only model in a probabilistic way. Thirdly, as an event progresses, firefighters interfere with natural fire growth (what models simulate) with suppression, altering natural fire growth. These issues hindered advances until new data enabled a new era in fire growth forecasting.

In simulations of events that occurred after the Suomi/NPP launch in 2011, a map of clearly delineated fire extent from the VIIRS instrument's active fire detection product could be used as initialization, to introduce the fire already in progress into a running CAWFE simulation, and also to evaluate simulations. Coen and Schroeder (2013) [9] presented a cycling forecasting approach, in which a sequence of CAWFE forecasts could be initialized using updated weather and VIIRS fire maps. In this approach, a sequence of simulations is initialized with new active fire detection data from VIIRS, and each simulation continued in time. As a new forecast becomes available, it is used as guidance and, the previous forecast, in which errors will have accumulated, is discarded. In principle, this allows fire growth to be simulated well for the next 1-3 days from the time the fire is first detected by satellite until it is extinguished.

In practice, both additional challenges and advantages arise [63]. For example, the cycling approach may provide a poor forecast if highly unpredictable weather occurs (for example, convective initiation) and the forecast deteriorates before another observation is available or if a fire observation is missing (e.g. the data collection is obscured by clouds, ruined by sun glint, or the viewing angle is too far off nadir). Those obstacles are balanced by the availability of new sensors the observations of which not only allow for asynchronous new forecasts to be added, perhaps improving the overall system skill by keeping the current forecast more up to date, but, given a detection closer to ignition, as in the Canyon Creek Wildfire [63], allow the forecast system to start simulations and guide response earlier.

CAWFE has also been configured as a deterministic forecast, in which it is fast enough to produce a forecast for a large fire (innermost domain 26 km x 26 km), for example, the Tubbs Fire, four times faster than real time on a single processor [23]. Simple prototype coupled systems, with simulations manually triggered at notification of a new fire, have been demonstrated in [64-67], where WRF-based systems [65-67] employ hundreds - thousands of processors. More recent applications have used CAWFE to downscale operational forecasts in October 2019 during forecasted Diablo and Santa Ana wind events in northern and southern California respectively, to see if it is possible to more precisely identify extreme wind locations [62] and refine PSPSs.

### 3.2.2 Results of applications to land management in complex environments

In forested mountain environments, fire is a natural disturbance and prescribed fire the tool of choice for reducing fuel load and restoring lands to a healthy condition. Fires in this environment are forged by intricate mountain winds, including winds amplified by the fire itself. Unplanned prescribed fire escapes have been costly, sometimes costing lives, reducing public acceptance of this mitigation technique despite awareness of a building wildfire potential. Only a small percentage of areas needing treatment are completed. In this environment, current operational fire modeling tools are particularly challenged, making fires appear unpredictable, increasing uncertainty, and inhibiting greater use of prescribed fires that might mitigate the hazard. A key application of coupled models is in investigating the potential for large fire growth in such complex environments

Fine-scale topographic effects on wind, microscale circulations, and fire-induced winds are less well appreciated among the forestry community than more tangible factors such as fuel loads and moisture, yet these may trigger branching of fire behavior into very different outcomes. Tarnay et al. (2018) [68] used CAWFE simulations to assess the risk of large fire potential from four hypothetical ignitions along the North Yuba River (Fig. 14) in Tahoe National Forest in the Sierra Nevada Mountains in the weak southwesterly wind conditions of Sept. 17, 2014 (the day on which the King Fire, 75 km to the south, ran 25 km (Section 3.1.1)). Ignitions #1, #2, and #4 produced small fires over the first 12 h burning period. Ignition #3, at the town of Goodyears Bar, CA, lay at the intersection of a north-south canyon and the river valley, allowing winds to penetrate down into the valley and fan the ignition. This nudged it into the inclined north-south canyon, where the plume it produced drew air in across the fire, beginning a fire-atmosphere feedback that caused the fire to grow over 12,100 ha during the same 12 h burning period. This highlighted the alarming potential for severe fires in conditions that do not themselves appear severe, such as the potential for megafires in the Sierras during common autumn weak southwesterly winds and the very different outcomes that could arise from ignitions in close proximity due to local topographic effects.

Computational models are uniquely able to show that in these complex environments, there can be greater variability than imagined in potential fire growth due to subtle differences in the fire environment or perhaps counterintuitive (in the case of previous beetle kill impacts on live fuel load [24]) results when considered in a dynamic framework.

*3.3  Transient phenomena: Drawing knowledge to enhance firefighter safety*

Firefighters are vulnerable to complex, transient dynamic fire phenomena as well as "perfect storm" scenarios where numerous fuel, local topographic wind effects, fire amplification as it is channeled though canyons, and feedbacks between the wind and fire come together. Albini (1984) [69] described the state of knowledge of fire behavior and fire phenomena at the time, emphasizing the uncertainty and some hypotheses about how they occurred, particularly transient phenomena. A notable example that still poses a safety hazard are fire vortices, which may range from the small meter-sized fire whirls along a fire line to firenadoes that spin off an actively burning fire potentially with damaging winds of tornadic strength, to situations where the entire plume over a fire is rotating and may resemble a meteorological supercell storm, with updrafts up to 50 m s$^{-1}$ and indrafts into the fire that may reach 40 m s$^{-1}$. Fire whirls themselves are identified as an example of extreme fire behavior indicating direct suppression should be stopped. Rotating plumes have proven to be safety problems, for example, during the 2008 Indians Fire, as the rotating convective column swept over a crew seeking shelter in and behind an engine, the flames curling around the vehicle. A similar event occurred during the 2002 Missionary Ridge Fire, as a firenado spun off the primary convective column across a dry lakebed that had been designated a safety zone. The Carr Fire, in which a confluence of downslope westerly winds intersected northerly flow up the Sacramento Valley, produced a shear zone across which a fire line with rapid consumption of heavy fuels combined to produce a firenado 5.5 km tall, 300 m wide, with winds exceeding 230 km h$^{-1}$ - a phenomena that threw a firefighter's truck over 500 m. Images of the Carr Fire tornado were analyzed by [71]. Pulsing or collapsing motions of the convective column over a fire are anecdotally connected to sudden amplification or other changes in fire behavior, though science has not yet unraveled this. Research is still exploring how, where, or how frequently such phenomena may occur on large wildland fires.

    Investigation of fire phenomena is at the forefront of computational modeling of fire behavior. Simulations have begun to capture more complex fire phenomena and illuminated the conditions in which they form, including the formation of fire whirls along the edge of a fire [45]. CAWFE simulations of the Esperanza Fire [7] produced a 1-km wide firenado on the upslope flank and horizontal roll vortices (HRVs) (pairs of counter-rotating updrafts laid forward on their side) that formed where northeasterly winds driving the fire met southwesterly winds at the fire line providing the horizontal wind gradient needed (when lifted by the fire convective column) to generate rotation around a vertical axis. These atmospheric phenomena impact fire behavior, as HRVs increased (decreased) the fire rate of spread in between vortices where

downdrafts (updrafts), respectively, coincided.

Some of the most hazardous and difficult to anticipate behavior arises when precipitation falling from a cloud evaporates as it falls, cooling the subcloud air, and generates a cool gust front that spreads outwards. These outflow boundaries can persist for 24 hours and travel over a hundred kilometers. When an outflow boundary encounters a fire, shifting winds can rapidly change the intensity and direction of spread. This is exemplified by the 2013 Yarnell Hill Fire, which resulted in 19 firefighter deaths. While current training materials emphasize the outflows from thunderstorms, small, nearly transparent clouds (known as "virga") may produce strong downdrafts and outflows as well. The Frog fire (1 fatality) was believed to have been influenced by the collapse of several small cells that produced multiple wind shifts. Some outflow winds may be generated by the pyrocumulus clouds formed by the fire itself. A possible example is the Las Conchas fire, which generated a deep, precipitating cloud over the fire. Early analysis suggests that precipitation falling from the cloud evaporated as it fell, cooling the subcloud air, and generated a gust front that without warning spread the fire downslope an additional 20,000 ha overnight when weather conditions suggested no growth would occur.

Gust front-driven fire behavior cases are some of the most difficult to reproduce with computational models due to the need to accurately represent the location and timing of a sequence of events: convection initiation, the onset, production, and descent of precipitation, gust front propagation, its impact with the fire, and subsequent fire behavior. Fig. 15 (adapted from [21]) shows the reported progression and the innermost computational domain of a CAWFE simulation of the Yarnell Hill Fire. On the day of the fatality, southwesterly winds crossed central Arizona, pushing an existing fire north, where it encountered a westerly gap flow, turning it east, as seen in the report's map of fire progression (Fig. 15, left). Simultaneously, a high-based tongue of moist air from the Pacific northwest had begun crossing a series of mountains in northeast Arizona, triggering thunderstorms. As their rain fell into a dry boundary layer, it evaporated, creating a gust front that spread to the southwest. The gust front encountered the fire, changing the direction it was spreading from slowly towards the east to rapidly toward the southwest, driving it over the sheltered firefighters. Subsequent investigation suggests a pyrocumulus over the fire may have contributed to additional strengthening of the fire (M. Fromm, pers. comm.) Though firefighters are given warnings that thunderstorm outflows of a particular strength are likely in an afternoon, specific warnings on particular features are not done. In part, this reflects the difficulty in predicting specifics of convective initiation, but also reflects the inability of current kinematic operational fire behavior models to simulate this sort of event. This event exemplifies where a coupled numerical weather prediction–fire behavior model can advance understanding and firefighter safety beyond current fire modeling tools, by linking the fire behavior, the cloud it generates, the precipitation that falls from the cloud, the gust front this produces, and the resulting feedback on fire behavior.

## 4. Conclusions

The recognition that much of the current complexity surrounding wildland fires arise from their interaction with the fluid surrounding them has led to an expansion in the use of computational models. These models' uses range from basic understanding of fire behavior, investigation into the causes of extreme events, more sophisticated forecasting systems based on perceptions that current kinematic models no longer suffice, to developing training material on this new understanding for firefighters. Here, we have synthesized recent studies using convective-scale (hundreds of meters horizontal grid spacing) simulations of the weather leading up to and early fire growth during recent wildfire events with the CAWFE® coupled weather-wildland fire modeling system, and others for comparison, using satellite active fire detection data from the Visible and Infrared Imaging Radiometer Suite (VIIRS) instrument and airborne fire mapping to initialize and evaluate event simulations. Taken as a whole, they provide insight on current wildland fire issues and areas of progress.

Computational science is bringing dramatic advances and technology to this important hazard. For example, it is slowly changing widely repeated, firmly believed misunderstanding about causes of why a fire became large and what factors are important. Firefighters speak of a dichotomy of "plume-driven" vs. "wind-driven" fire behavior; most extreme events frequently lie at each end of this spectrum. However, studies of events like the 2006 Esperanza Fire or the 2017 Thomas Fire have shown a single fire may have both wind- and plume-driven components, the location and timing of which vary throughout an event due to changing of synoptic pressure pattern, how winds are oriented with respect to topography, and how topography alters, blocks, or shelters certain areas. During a wind-driven event, components of the fire can spread rapidly orthogonal to wind - a development that in those cases led to firefighter fatalities. The largest or most destructive fires can occur due to or along with exceptional atmospheric extremes (e.g. during California or Colorado wind events), others do not require extreme conditions, and arise from a fortuitous alignment of small, local factors. Fire behavior may yet surprise due to the lack of apparent ambient triggers visible in surface weather station network data or coarse operational NWP model forecasts - most often during "plume-driven" fires due to fire-induced winds arising from internal dynamics or aided by undetectable microscale circulations occurring between sparsely located stations or model grid points. Computational science studies have allowed the testing of "what if" scenarios, testing the impact of various factors on an event, sometimes providing counterintuitive answers, for example, that despite occurring within a drought or in forests experiencing built up fuels, these factors, though dead fuel moisture and fuel load effects, have secondary effect except on fires climbing inclined especially concave terrain slopes, where the factors can reinforce each other through the atmospheric medium.

In plume-driven events, studies showed that the actual potential for large fire growth may be much greater or less (depending on the location) than previous studies using kinematic models indicated. "Extreme" fires can occur during conditions (weather) that are not extreme and the primary factors shaping the fire and driving the rapid growth (microscale circulations and fire-induced winds) may not be apparent. Other factors (such as fuel moisture and loads) may only have noticeable impact where fires are growing upslope, where the factors may reinforce each other. Microscale factors such as small topographic features and aspects due to chance such as ignition location with respect to topographic features can make or break large fire growth in otherwise similar conditions through internal dynamic interactions.

Mesoscale model simulations - retrospective research and operational NWP forecasts both often based on the widely used community WRF modeling system - of the weather leading up to and during these events reproduce or anticipate, respectively, the occurrence of a wind event, capture broad spatial patterns of accelerated winds, and the degree to which the event is unusual. In and of themselves, they have underestimated the peak winds by 5-10 m s$^{-1}$ in their direct simulations, though users claim success by augmenting outcomes using "gust factors" to reach desired results. Assessing simulation-based claims of peak winds using longstanding NWS and RAWS surface station networks has been challenging, as complex terrain creates high spatial and temporal variability, and their placement was not targeted toward high wind extrema locations. (This perhaps increases the challenge for emerging machine learning approaches, in that though wind is the most critical environmental factor driving fire growth, wind data capture neither the fine-scale topographic flows nor fire-induced winds that drive fire behavior.) However, in several cases, ignitions and rapid early fire growth appear linked to strong winds and local extrema. Along with destruction, anecdotal evidence, and surface weather stations placed in California by utility companies in areas focused on wind corridors, have since detected higher gusts reaching 30-45 m s$^{-1}$.

However, when the operational NWP forecasts are used as initial and boundary conditions and downscaled with CAWFE, simulations generate microscale phenomena and wind extrema comparable to newer utility weather stations and distinctive features and characteristics of fire growth, including both wind-driven and fire-induced wind plume-driven components. While these wind features are small, sub-mesoscale circulations, they are not weak nor merely 'gusts', but resolvable dynamic phenomena with extreme wind speed peaks, often located near ignitions attributed by investigative reports to utility equipment faults. Microscale weather-fire behavior simulations of recent California wildfire events including the Tubbs Fire, the Redwood Valley Fire, the Thomas Fire, and the Camp Fire have revealed that (1) within regional wind events, exceptional narrow rivers or streaks of high wind speeds may lie between mesonet stations, producing wind extrema reaching 30-40 m s$^{-1}$, (2) certain locations are prone to conditions where wind speed and direction, atmospheric stability, and topography profile combine to generate locales of extreme, gusty winds that overlap utility and public assets. Studies found common factors such as a shallow (1-1.5 km) river of fast-moving stable air but several different types of dynamic microscale flow regimes, many of which produce dynamic microscale airflow regimes not present in the scientific literature and wind extrema that are co-located with the ignition area of highly destructive wildfires. Satisfying validation techniques are still a challenge, as standard observations (surface weather station networks, mapping of fires from remote sensing) can mislead or be misinterpreted as well as inform in these conditions.

Significantly, these flow regimes, phenomena, extreme wind speeds, and subsequent fire behavior have not been reproduced by WRF-based coupled models. While in some cases, this variance has been obscured by the unmentioned use of ad hoc "calibration" factors to increase or decrease spread rate to better match observations, in other events - some of the most destructive - simulations failed to capture the flow regime and if a prediction, would have been that destroyed towns would be safe from the fire. Although many models can be downloaded and used as a black box, this reminds of the need for deep subject matter understanding when using complex models. Understanding the magnitude, structure, and spatial distribution of these phenomena and wind extrema is currently one of the most pressing, highest economic impact atmospheric and computational science issues of today and has significant consequences for infrastructure design, operation, and preventing damage from additional events.

**Acknowledgements**

NCAR is sponsored by the National Science Foundation (NSF). This material is based upon work supported by FEMA under Award EMW-2015-FP-00888 and NSF under Award 1561093. Any opinions, findings, and conclusions or recommendations expressed in this material are the authors' and do not reflect the views of NSF.

26. C. C. Simpson, J. J. Sharples, and J. P. Evans. Resolving vorticity-driven lateral fire spread using the WRF-Fire coupled atmosphere-fire numerical model. Nat. Hazards Earth Syst. Sci. 14 (2014) 2359–2371.

27. T. L. Clark, W. D. Hall, J. L. Coen. Source Code Documentation for the Clark-Hall Cloud-scale Model Code Version G3CH01. NCAR Technical Note NCAR/TN-426+STR (1996) doi:10.5065/D67W694V.

28. T. L. Clark, T. Keller, J. Coen, P. Neilley, H. Hsu, W. D. Hall. Terrain-induced Turbulence over Lantau Island: 7 June 1994 Tropical Storm Russ Case Study. J. Atmos. Sci. 54 (1997) 1795-1814.

29. R. C. Rothermel. A Mathematical Model for Predicting Fire Spread in Wildland Fuels. Research Paper INT-115; USDA Forest Service, Intermountain Forest and Range Experiment Station: Ogden, UT. (1992)

30. F. A. Albini. PROGRAM BURNUP: a simulation model of the burning of large woody natural fuels. Final Report on Research Grant INT-92754-GR by USDA Forest Service to Montana State University, Mechanical Engineering Department, Bozeman, Montana, USA. (1994)

31. R. C. Rothermel. Predicting behavior and size of crown fires in the Northern Rocky Mountains. Res. Paper INT-438. Ogden, UT: U.S. Department of Agriculture, Forest Service, Intermountain Forest and Range Experiment Station, 46 p. (1991)

32. E. N. Stavros, J. Coen, B. Peterson, H. Singh, K. Kennedy, C. Ramirez, D. Schimel. Use of imaging spectroscopy and LIDAR to characterize fuels for fire behavior prediction. Remote Sensing Applications: Society and Environment 11 (2018) 41-50. https://doi.org/10.1016/j.rsase.2018.04.010
44

Table 1. Wildfires discussed here, along with their year of occurrence and location.

| Fires | Section | Figure | Year of occurrence | Location | Notes |
|---|---|---|---|---|---|
| High Park | 2.1.1, 3.1.1, 3.1.2 | 3 | 2012 | CO | Downslope wind-driven fire |
| King | 2.1.2, 3.1.1, 3.2.2 | 2 | 2014 | CA | Plume-driven fire |
| Rim | 3.1.1 | 1 | 2013 | CA | Plume-driven fire |
| Witch/Guejito | 3.1.2 |  | 2007 | So. CA | Santa Ana-driven fire |
| North Bay fires (Tubbs, Redwood Valley) | 3.1.2, 3.2.1, 4 | 4, 5, 6 | 2017 | No. CA | Diablo wind-driven fires |
| Thomas | 3.1.2, 3.1.3, 4 | 11, 12, 13 | 2017 | So. CA | Santa Ana-driven fire, plume-driven periods and areas. Firefighter fatality. |
| Camp | 3.1.2, 4 | 7, 8 | 2018 | CA | Diablo wind-driven fire |
| Chimney Tops II | 3.1.2 | 9, 10 | 2016 | TN | Downslope wind-driven fire |
| Esperanza | 3.1.3, 3.3, 4 |  | 2006 | So. CA | Santa Ana-driven fire, plume-driven components. Firefighter fatalities. |
| Canyon Creek | 3.2.1 |  | 2015 | OR | Lightning ignition, frontal passage |
| October 2019 CA wind/wildfire events | 3.2.1 |  | 2019 | No. CA, So. CA | Diablo wind-driven fires, Santa Ana-driven fires |
| Ignitions Tahoe Natl. Forest | 3.2.2 | 14 | 2013 | CA | Hypothetical ignitions in the Sierra Nevada Mountains |
| Missionary Ridge | 3.3 |  | 2002 | CO | Large fire whirl |
| Carr | 3.3 |  |  |  | Large fire whirl. Firefighter fatality. |
| Yarnell Hill | 3.3 | 15 | 2013 | AZ | Gust front-driven fire. Firefighter fatalities. |
| Frog | 3.3 |  | 2015 | No. CA | Gust front-driven fire. Firefighter fatality. |
| Las Conchas | 3.3 |  | 2011 | NM | Possible gust front-driven fire. |

**Figure captions.**

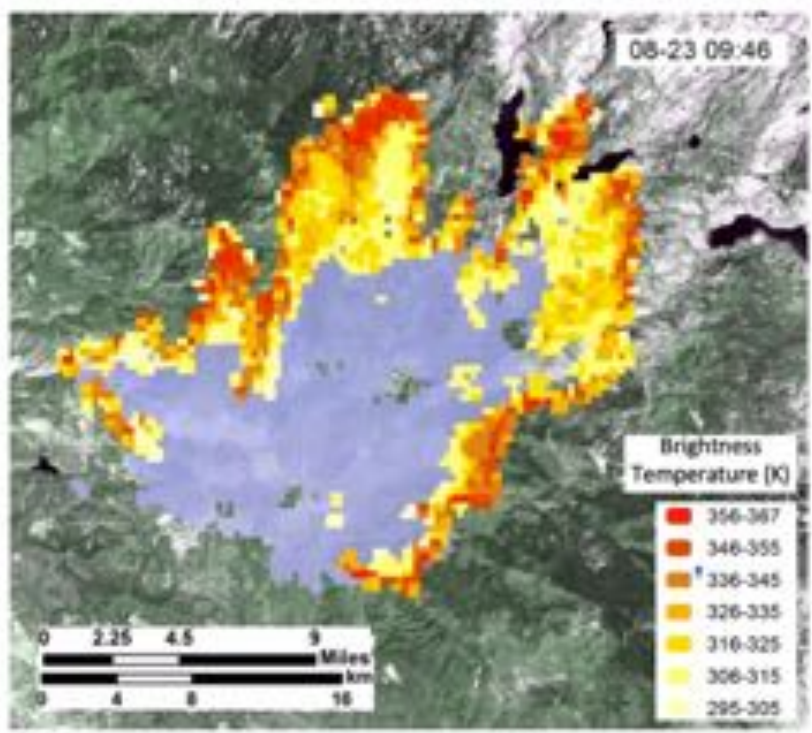

Figure 1. The Rim fire imaged in the medium wave infrared on Aug. 23 2013 by VIIRS.

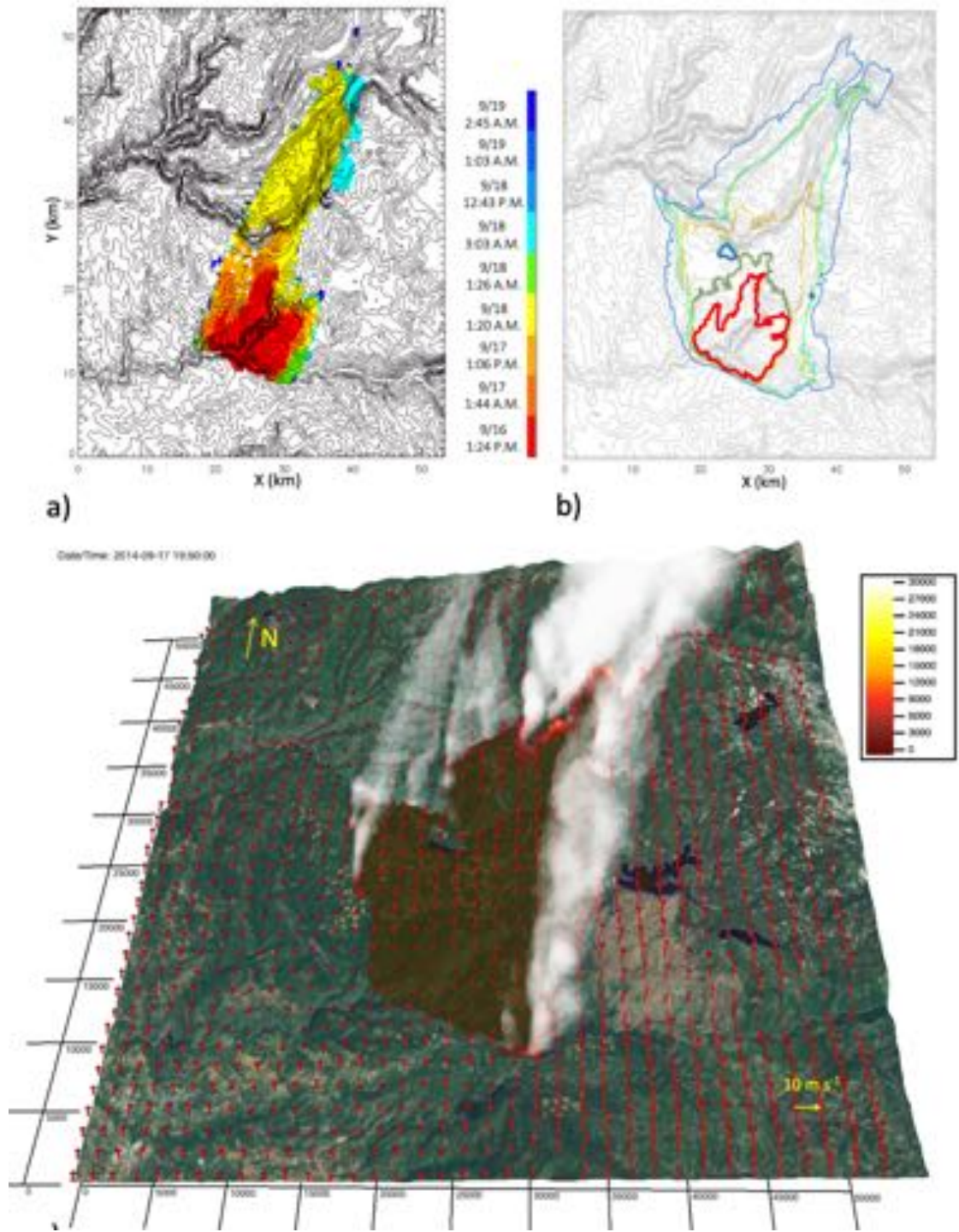

Figure 2. Observed and simulated fire progression during the 2014 King Fire. (a) NIROPS fire extent used to initialize fire in progress (red) (9:49 PM on 16 Sept.) and later VIIRS through 2:45 AM on 19 Sept. (b) Simulated fire extents at VIIRS detection times to 12:43 PM on 18 Sept., plotted with corresponding colors. (c) Simulated heat flux (in W m-2, according to color bar at right), smoke concentration, and near surface wind vectors at 7:50 PM on 17 Sept. Reprinted from [8] with permission from the Ecological Society of America.

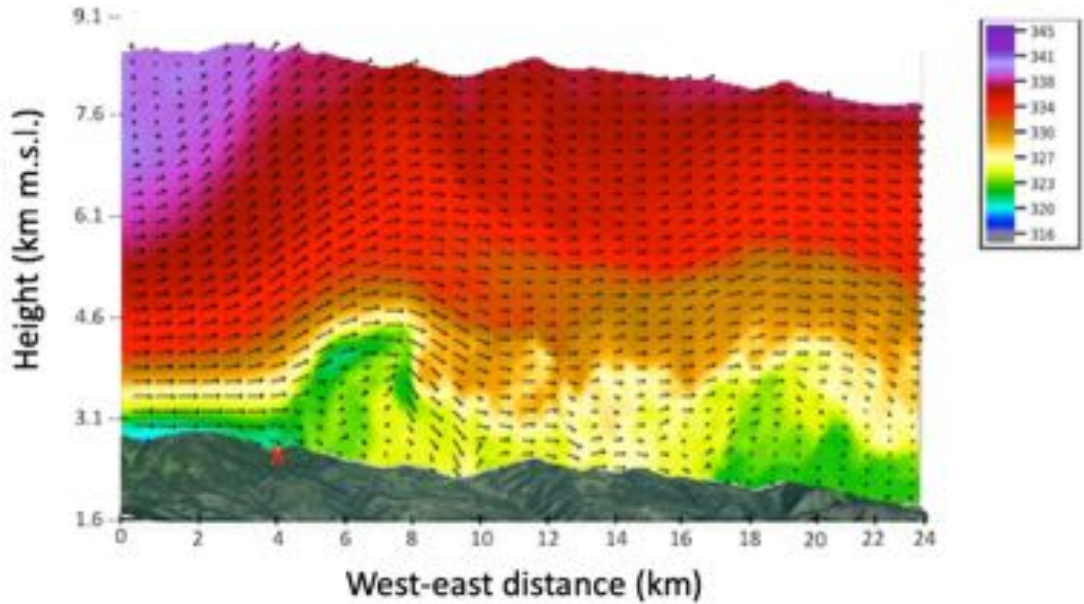

Figure 3. West-east cross section of potential temperature through an overturning gravity wave breaking over the High Park Fire. Reprinted from [24] with permission from the American Geophysical Union.

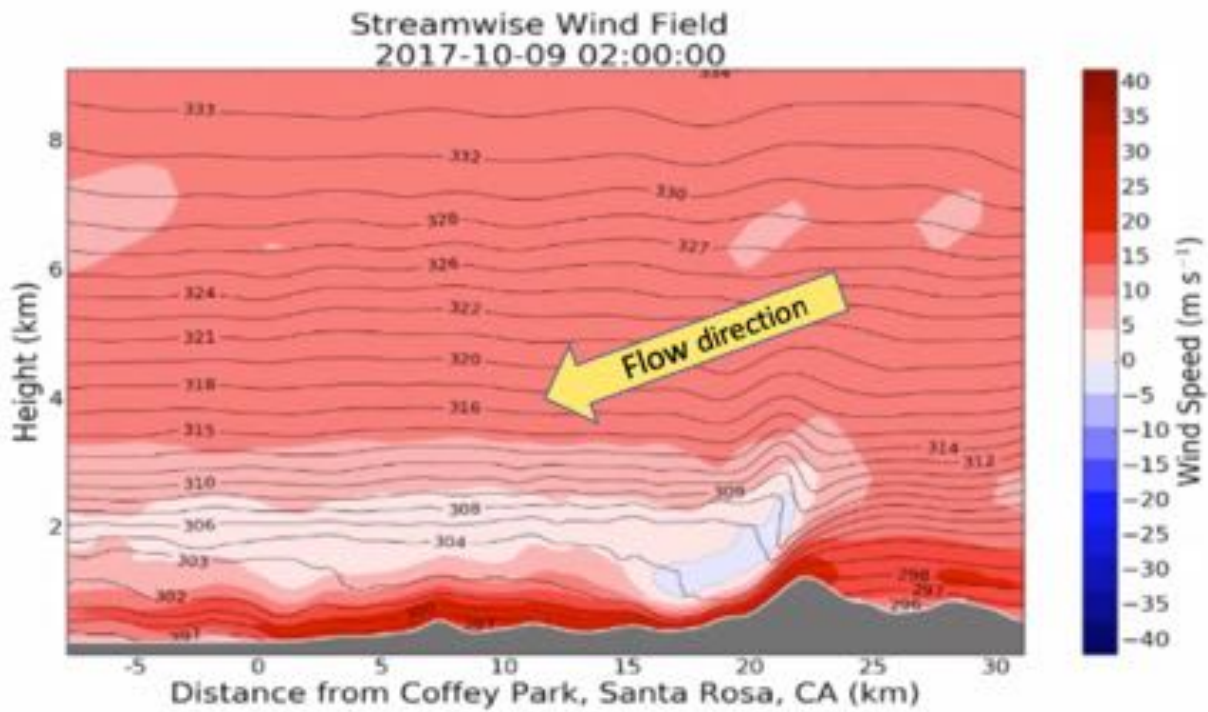

Figure 4. Vertical cross section of potential temperature, the vertical gradient of which indicates atmospheric stability, and streamwise winds (scale at right) through WRF weather simulation, 300 m horizontal grid spacing, during the Tubbs Fire. Reproduced from [57].

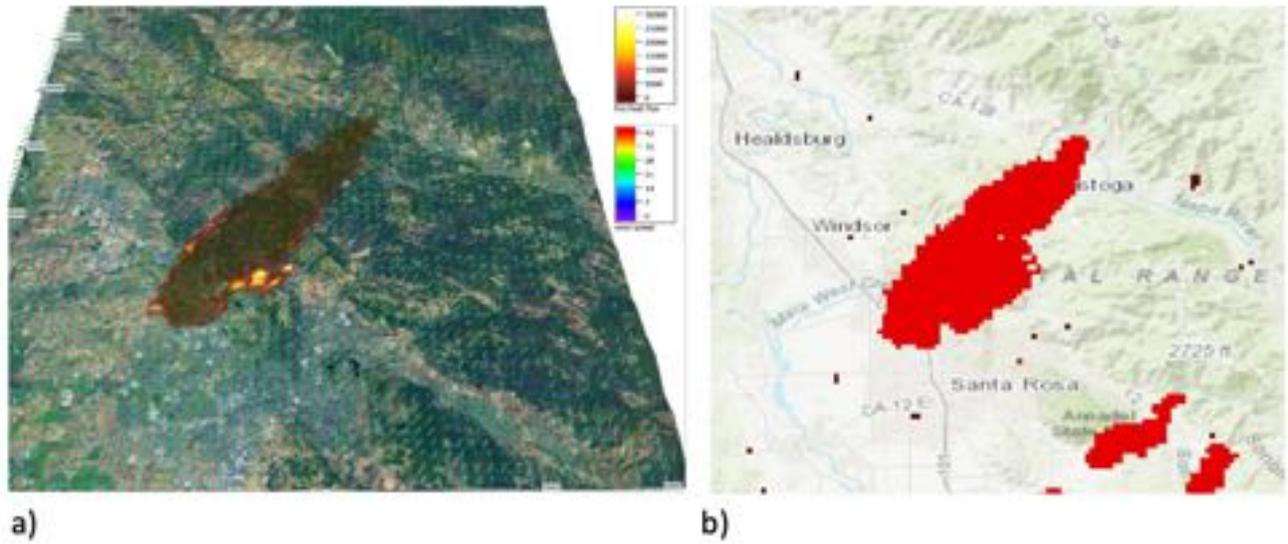

Figure 5. (a) CAWFE simulation of the Tubbs fire at 4:09 A.M. showing the downwind run of over 12 miles into the town of Santa Rosa in approximately 3 hours. The protrusion on the lower right side, orthogonal to the wind, occurred as the flank of the fire drew itself up a topographic feature. (b) Visible and Infrared Imaging Radiometer Suite (VIIRS) active fire detections at 3:09 A.M. Oct 9. 2017. Reprinted from [23].

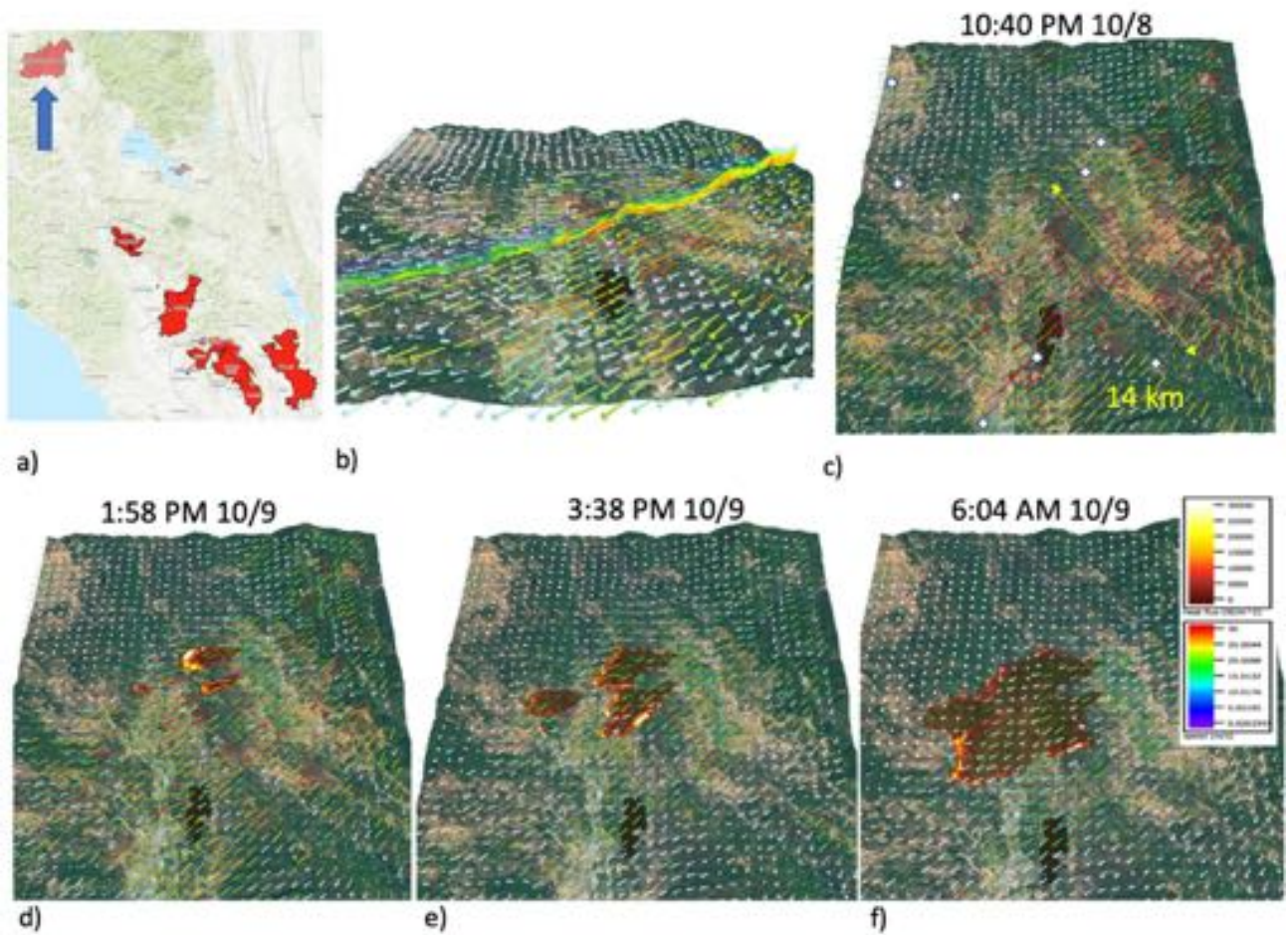

Figure 6. (a) Location of the Redwood Valley Fire amidst the October 2017 North Bay fire events. (b) High pressure inland drove air down-gradient over a lower barrier in the Sierra Nevada Mountains, creating a shallow, narrow river of high speed air that reportedly ignited the Redwood Valley Fire. (c) CAWFE simulation of the near-surface winds (vectors, colored according to color bar at right) and fire growth. Surface weather stations, only two of which were brushed by the river of strong winds, are indicated by circles. (d)-(f) CAWFE simulation at later times, where color contours indicate the fire's heat flux, colored according to color bar at right.

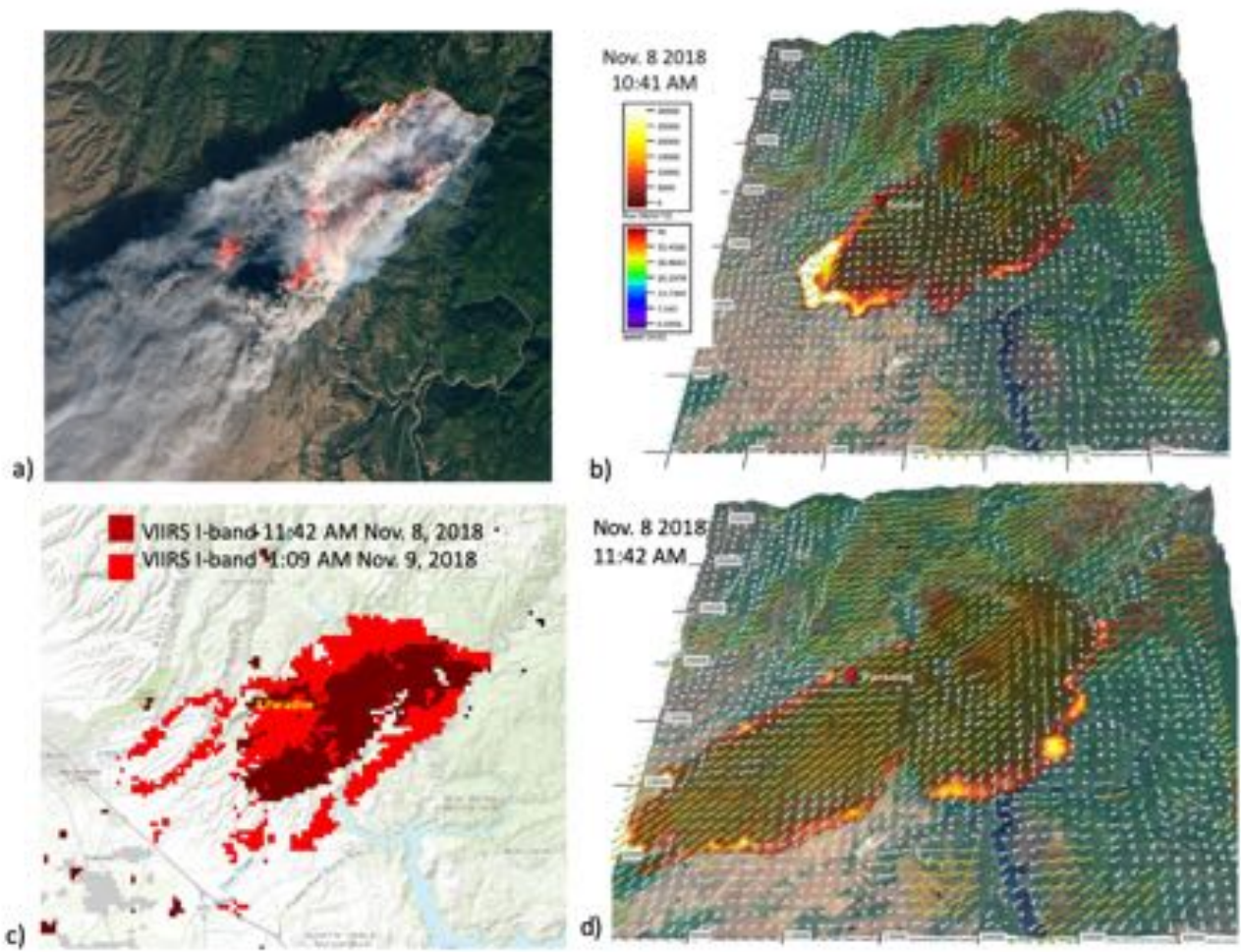

Figure 7. (a) A natural color image using visible and shortwave infrared bands to highlight the active fire using data acquired by the Operational Land Imager on Landsat 8 on Nov. 8 at 10:45 a.m. (b) VAPOR visualization of a CAWFE® coupled weather-wildland fire model simulation of the November 8, 2018, Camp Fire in Paradise, CA (indicated in figure) at 10:41 a.m. The image shows the heat flux produced by the fire, colored according to the color scale at left (top). The arrows indicate the simulation wind speed near the surface, colored by the wind speed magnitude (according to lower color bar at left). (c) Extent of the Camp Fire as mapped by VIIRS at 11:42 AM (colored brick) on Nov 8 and near the end of the first day of growth (colored red). (d) extent at 11:42 a.m. as simulated by CAWFE (comparable to brick-colored shape in (c)).

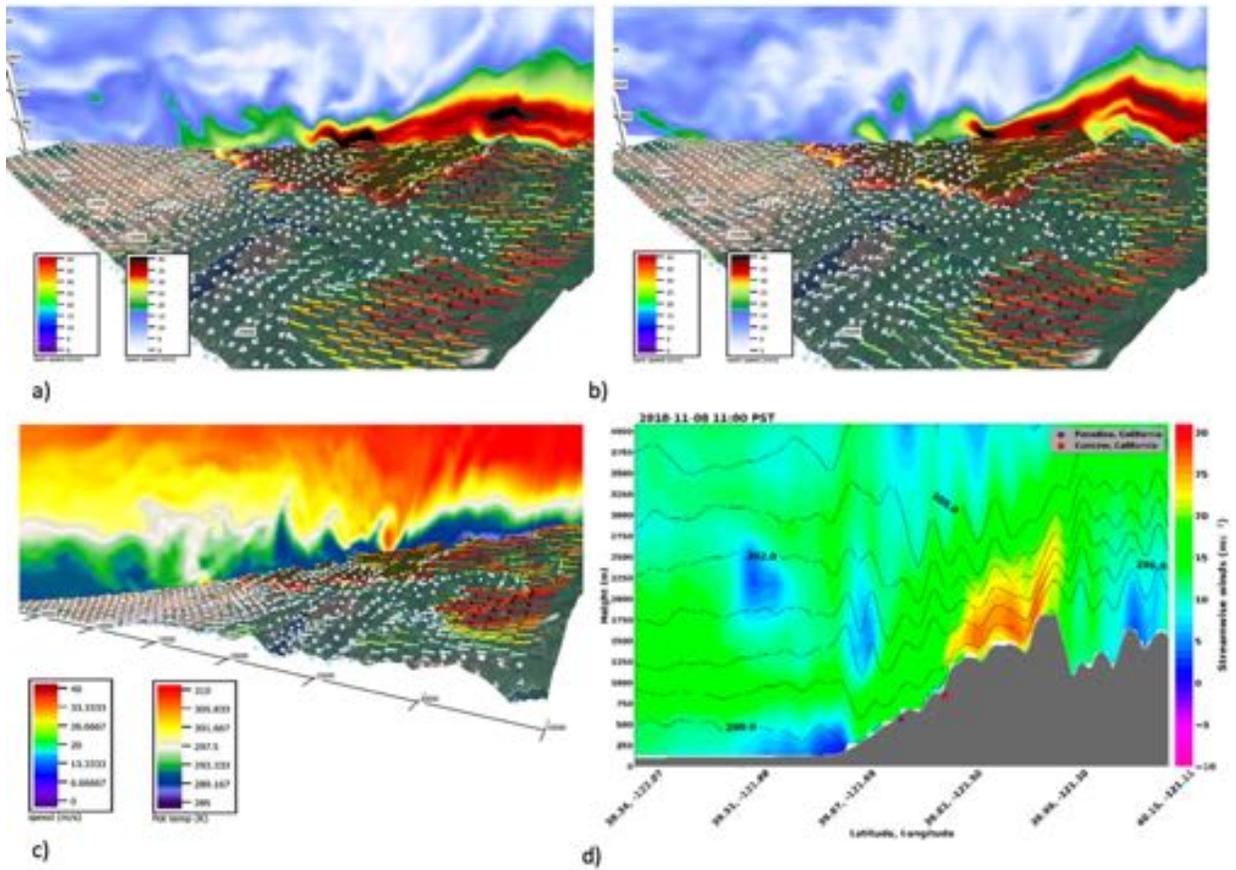

Figure 8. Vertical cross section through the center of the Camp Fire simulated with CAWFE showing (all colored according to color bars at right) (a) the wind speed in the plane at 10:17 AM and near surface wind vectors colored according to color bar at lower left, (b) the wind speed in the plane at 10:27 AM, (c) the atmospheric potential temperature, and (d) from a WRF simulation of the weather, reproduced from [59], the potential temperature and streamwise winds.

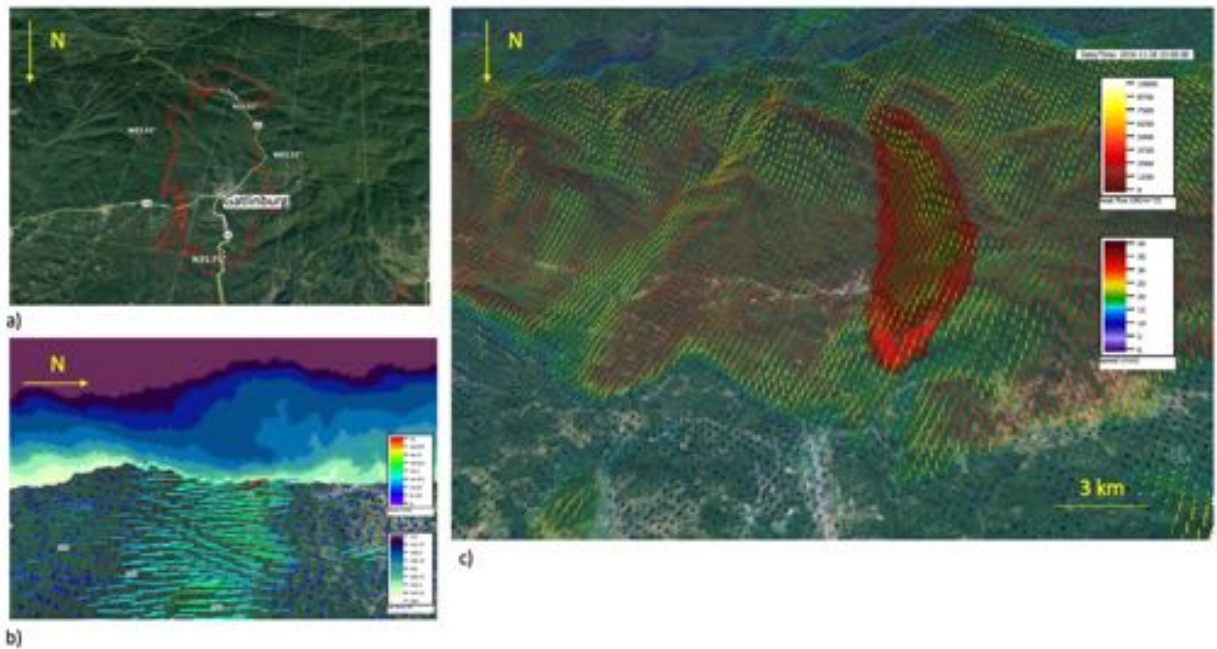

Figure 9. (a) Extent of the Chimney Tops 2 as mapped by Fire National Infrared Operations aircraft at 7:40 PM on Nov. 29, 2016. (b) a vertical cross section through a CAWFE simulation of the event, taken longitudinally down the center of (a), showing the near surface downslope winds and jump-like structure in the atmospheric potential temperature, and (c) plan view of the fire extent at 11 PM Nov. 28, 2016 simulated by CAWFE. Near surface wind vectors shown the extent of the strong downslope winds.

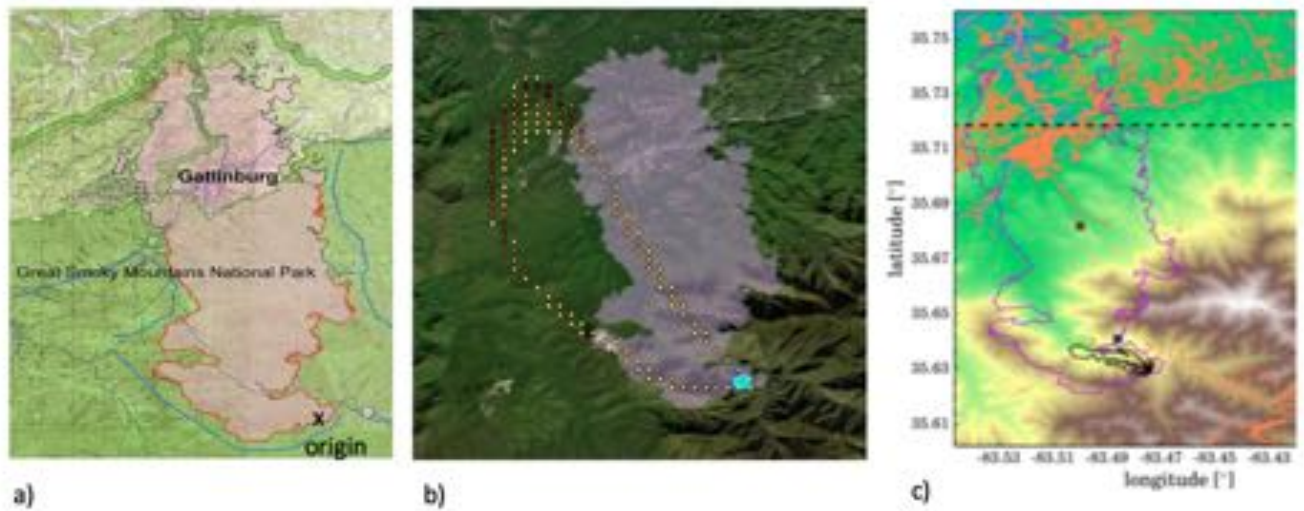

Figure 10. The devastating growth period of the Chimney Tops 2 Fire, (a) as mapped afterwards by National Infrared Operations aircraft at 7:40 PM on Nov. 29, 2016, (b) as simulated by CAWFE until 11 PM on Nov. 28, after which winds subsided, where dots indicate actively burning regions of the fire, overlaid on the outline from (a), and (c) the simulation (black contour) using a WRF-based coupled modeling system, WRF-FIRE, reproduced from [60].

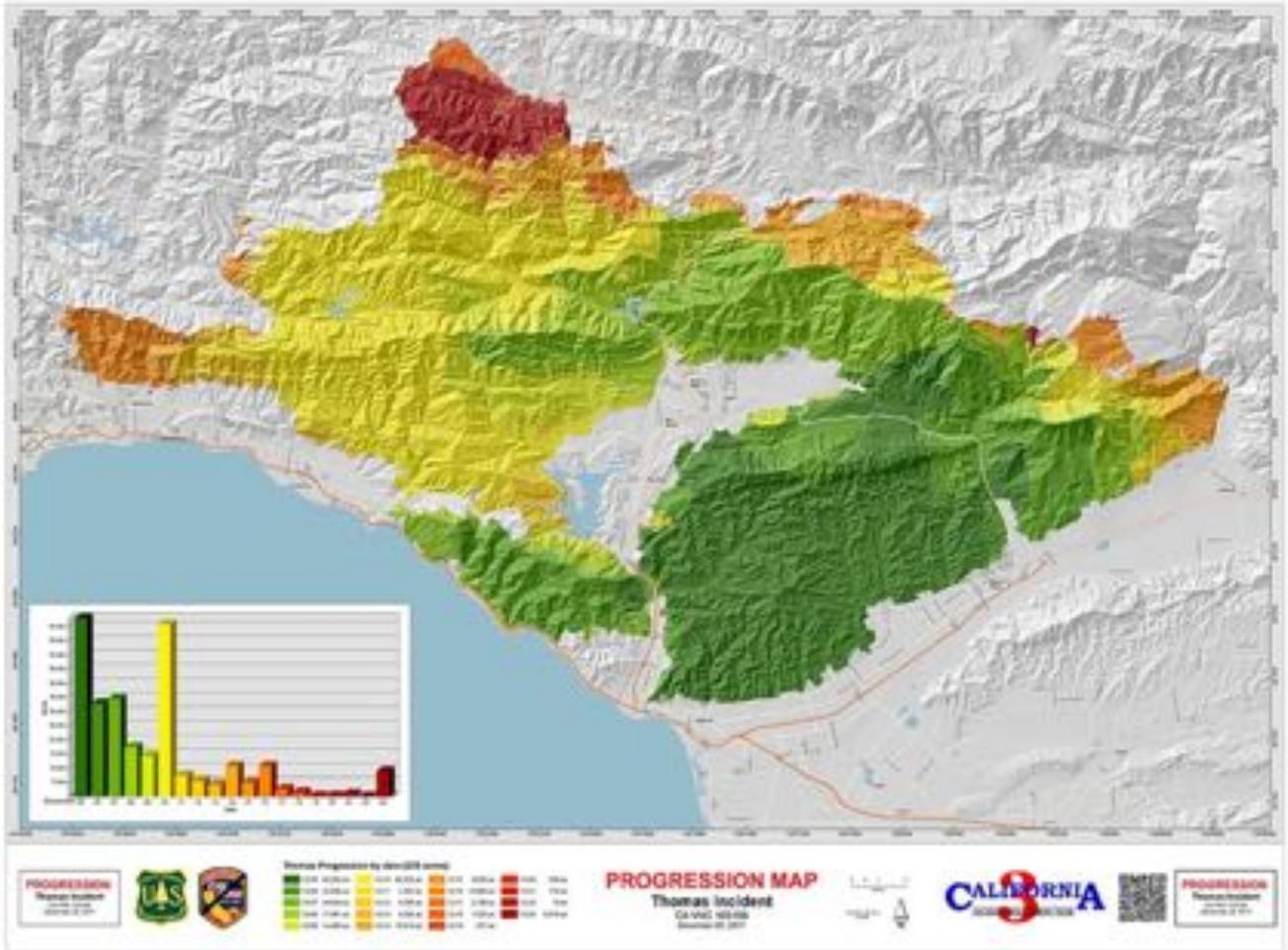

Figure 11. Progression of the 2017 Thomas Fire. Reprinted from California Department of Forestry and Fire Protection.

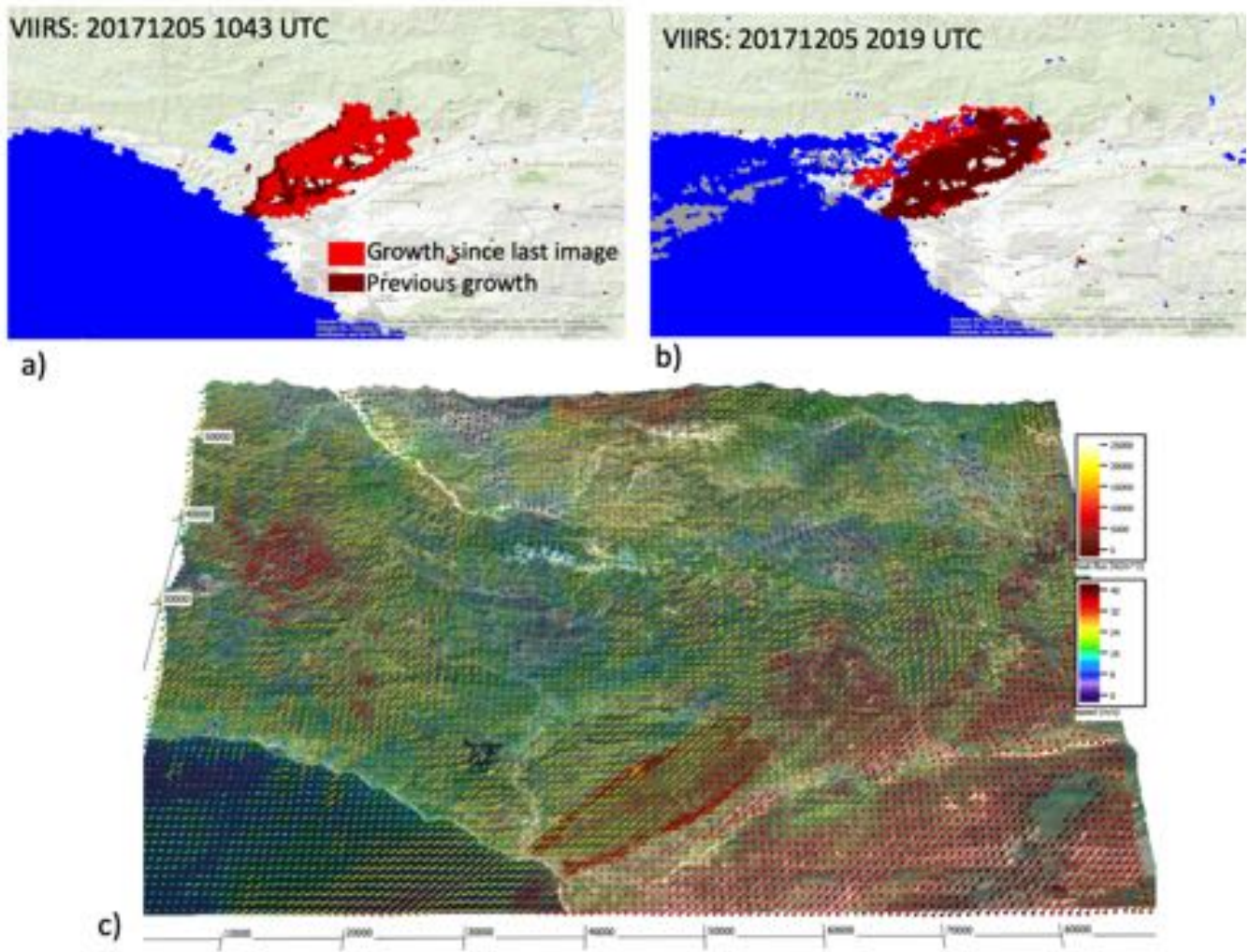

Figure 12. Thomas Fire extent during the early wind-driven period showing growth since the last image (red) and previous extent (brick) VIIRS active fire detections a) at 2:43 a.m. on 5 Dec. 2017 and b) at 2:19 PM on 5 Dec 2017. c) CAWFE simulation of near surface winds during this period (arrows, colored by wind speed, according to scale at right, with faster wind speed in hotter colors) and fire heat flux (according to top scale at right).

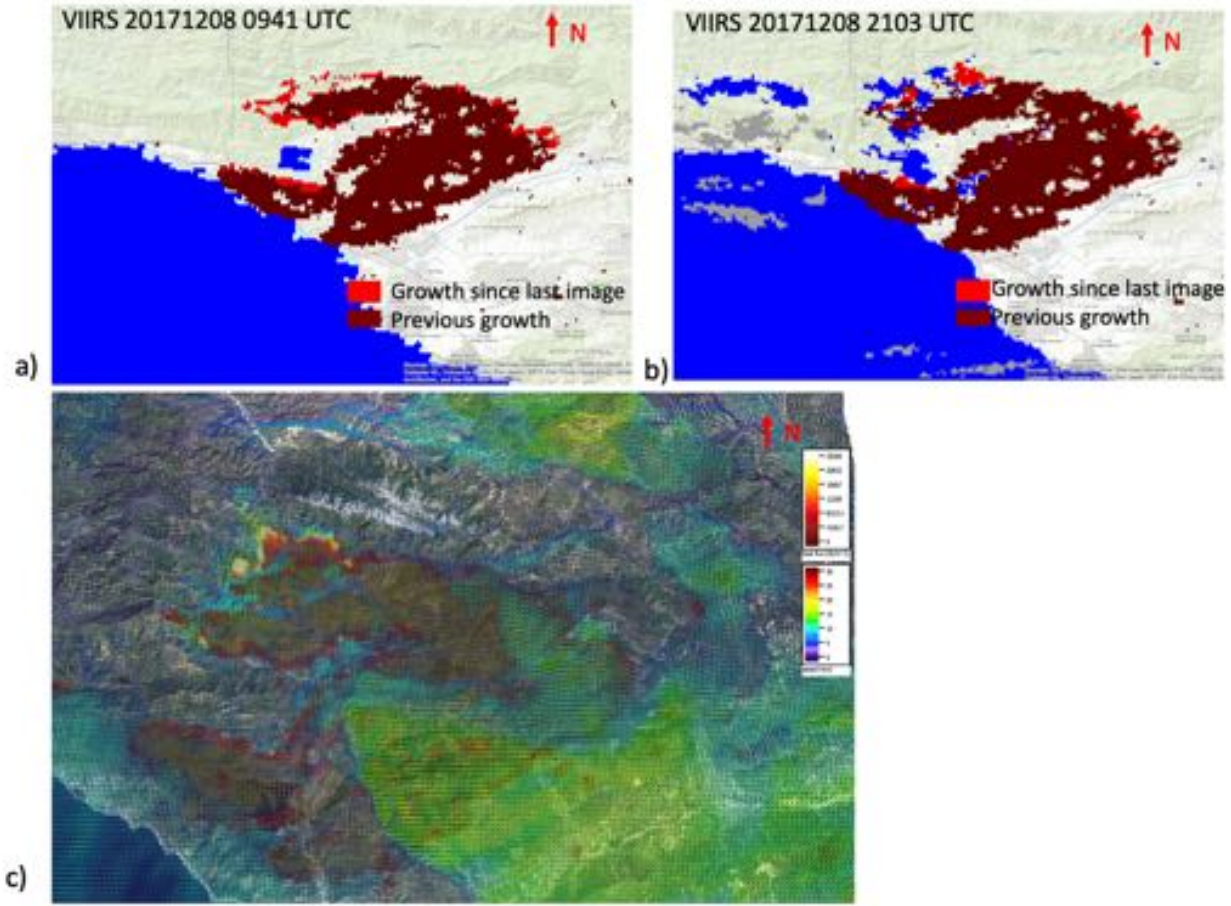

Figure 13. Thomas Fire extent and active growth areas during a later plume-driven period. a) Cumulative fire affected area at 1:41 a.m. PST on Dec. 8 (red) and previously detected extent (brick) from VIIRS active fire detections. b) Growth since the last image (red) and previous extent (brick) VIIRS active fire detections at 1:03 p.m. on 8 Dec. 2017. c) CAWFE simulation of near surface winds at 2:48 p.m. PST (arrows, colored by wind speed, according to scale at right, with faster wind speed in hotter colors), where Santa Ana winds continue but not on active growth areas of the fire, and fire heat flux (according to top scale at right).

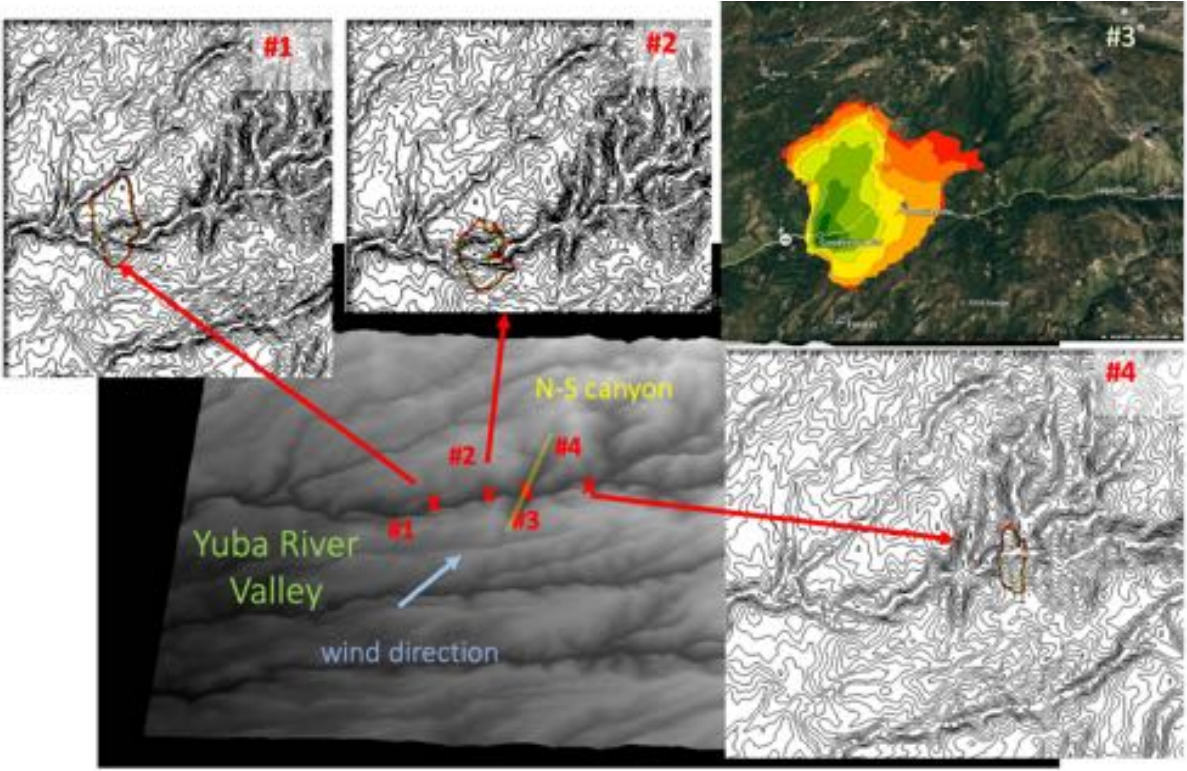

Figure 14. CAWFE simulations of the evolution of 4 hypothetical ignitions along the Yuba River Valley through the day's burning period in weather of 18 Sept., 2014, the day of the nearby King Fire (Fig. 2).

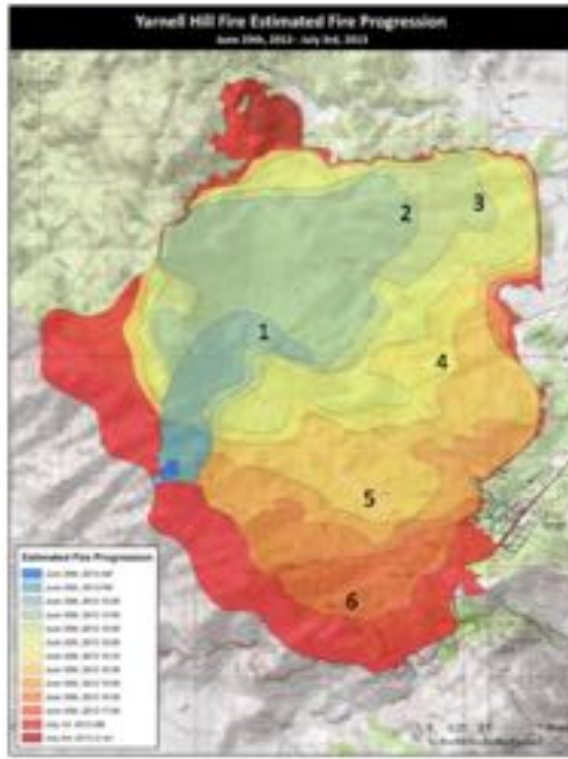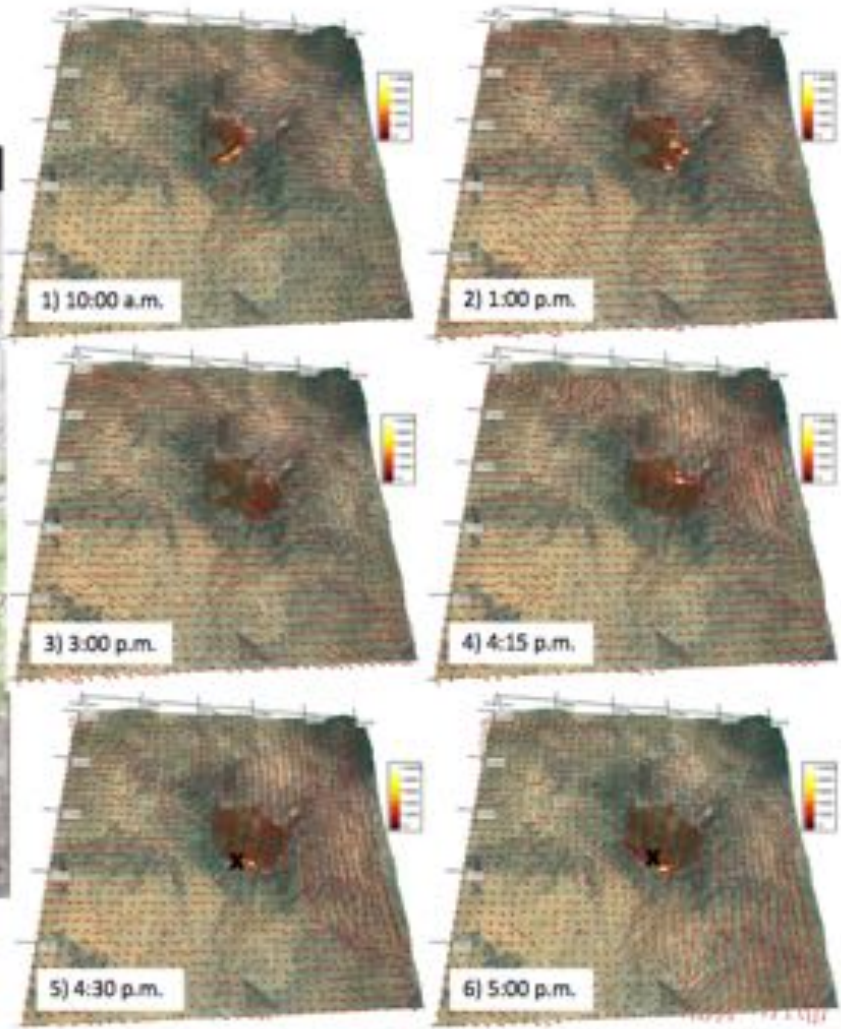

Figure 15. Yarnell fire progression from Serious Accident Investigation Report (left) and a sequence of times during a CAWFE simulation (numbered 1 to 6, right). Reprinted from [21].